\documentclass[aps,prl,reprint,superscriptaddress,nobalancelastpage,twocolumn,showpacs]{revtex4-1}
\usepackage{color,amsthm,amsmath,amsfonts,graphicx,bm,amssymb,hyperref, tikz, braket}
\usepackage{epstopdf}
\usepackage{comment}
\usepackage{etoolbox}
\usepackage{marginnote}
\newcommand{\titleinfo}{Inhomogeneous quenches in a fermionic chain: exact results}

\DeclareMathOperator{\tr}{tr}
\DeclareMathOperator{\ad}{ad}
\begin{document}
\newcommand{\beq}{\begin{equation}}
\newcommand{\eeq}{\end{equation}}
\newcommand{\Ord}[1]{{O}\left(#1\right)}
\newcommand{\sign}{\,{\rm sign}}
\newcommand{\arctanh}{\,{\rm arctanh}}
\newcommand{\subscript}[2]{$#1 _ #2$}

\providecommand{\abs}[1]{\lvert#1\rvert}  
\providecommand{\norm}[1]{\lVert#1\rVert}

\newcommand{\twovec}[2]{\left[\begin{array}{c}
#1\\#2
\end{array}
\right]}

\newcommand{\avg}[2]{\langle #1 \rangle_{\mbox{\tiny #2}}}
\def\Tr{\operatorname{Tr}}
\def\diag{\operatorname{diag}}

\def\frl{\phi_{r/l}}
\def\fr{\phi_{r}}
\def\fl{\phi_{l}}
\def\fRL{\psi_{R/L}}
\def\fR{\psi_{R}}
\def\fL{\psi_{L}}

\def\irl{\Phi_{r/l}}
\def\ir{\Phi_{r}}
\def\il{\Phi_{l}}
\def\iRL{\Psi_{R/L}}
\def\iLR{\Psi_{L/R}}
\def\tiRL{\tilde\Psi_{R/L}}
\def\iR{\Psi_{R}}
\def\iL{\Psi_{L}}
\newcommand{\ialpha}[1]{\Psi_{#1}}

\def\fLW{\omega}
\def\fFO{\Psi}
\def\fFW{\psi}
\def\fFOstat{\vec{\fFO}^{\mbox{\tiny stat}}}
\def\FRL{\vec{\Upsilon}}
\def\ttheta{\tilde\theta}

\def\rightP{\mathcal{P}}
\def\mysmall{\delta}

\def\MM{M}
\def\bigM{\mathcal{M}}
\def\bigN{\mathcal{N}}
\def\bigc{\mathbf{c}}
\def\bigQ{\mathcal{Q}}
\def\bigpsi{\mathbf{\Psi}}
\def\bigphi{\mathbf{v}}
\def\bigA{\mathcal{A}}
\def\bigB{\mathcal{B}}
\def\bigAt{\tilde{\mathcal{A}}}
\def\smallM{m}
\def\Hleft{H_l}
\def\Hright{H_r}
\def\smallh{h}
\def\betaleft{\beta_l}
\def\betaright{\beta_r}
\newcommand\pv[1]{P_v\left[#1\right]}
\def\mom{p}
\def\Li{\operatorname{Li}}
\def\rhostat{\hat\rho_{\text{\tiny stat}}}
\def\citare{{\color{red}[??]}}
\def\hH{\hat{H}}
\def\hO{\hat{\mathcal{O}}}
\def\Trev{T_{\text{rev}}} 
\def\Tth{T_{\text{th}}} 
\graphicspath{ {../} }
\def\partA{\mu} 
\def\partB{\nu} 

\newcommand\makered[1]{{\color{red} #1}}
\newcommand\makeblue[1]{{\color{blue} #1}}
\newcommand\repl[2]{\sout{#1}\makered{#2}}
\newcommand\mrepl[2]{\text{\sout{\ensuremath{#1}}}\makered{#2}}
\newcommand\note[1]{\marginnote{\tiny \textcolor{blue}{#1}}}

\newcommand\leftnote[1]{\reversemarginpar\marginnote{\tiny \textcolor{blue}{#1}}}
\newcommand\rightnote[1]{\normalmarginpar\marginnote{\tiny \textcolor{blue}{#1}}}
\def\vmax{v_{\mbox{\tiny max}}}

\title{Inhomogeneous quenches in a fermionic chain: exact results}

\author{Jacopo Viti}
\affiliation{Max Planck Institute for the Physics of Complex Systems, N\"othnitzer Str.~38, 01187
  Dresden, Germany}
\author{Jean-Marie St\'ephan}
\affiliation{Max Planck Institute for the Physics of Complex Systems, N\"othnitzer Str.~38, 01187
  Dresden, Germany}
\author{J\'er\^ome Dubail}
\affiliation{Groupe de Physique Statistique, IJL, CNRS/UMR 7198, 
Universit\'e de Lorraine, BP 70239, F-54506 Vand{\oe}uvre-l\`es-Nancy Cedex, France}
\author{Masudul Haque}

\affiliation{Max Planck Institute for the Physics of Complex Systems, N\"othnitzer Str.~38, 01187
  Dresden, Germany}

\begin{abstract}

  We consider the non-equilibrium physics induced by joining together two tight binding fermionic
  chains to form a single chain. Before being joined, each chain is in a many-fermion ground
  state. The fillings (densities) in the two chains might be the same or different.  We present a
  number of exact results for the correlation functions in the non-interacting case.  We present a
  short-time expansion, which can sometimes be fully resummed, and which reproduces the so-called
  `light cone' effect or wavefront behavior of the correlators.  For large times, we show how all
  interesting physical regimes may be obtained by stationary phase approximation techniques. In
  particular, we derive semiclassical formulas in the case when both time and positions are large,
  and show that these are exact in the thermodynamic limit.  We present subleading corrections to
  the large-time behavior, including the corrections near the edges of the wavefront.  We also
  provide results for the return probability or Loschmidt echo.  In the maximally inhomogeneous
  limit, we prove that it is exactly gaussian at all times.  The effects of interactions on the Loschmidt echo are also
  discussed.

\end{abstract}



\maketitle

\paragraph*{Introduction --} 
Local quantum quenches are a particularly neat setup to address theoretical questions about
non-equilibrium current-carrying stationary states as well as transport properties of many-body
isolated quantum systems.  The characterization of the long-time behavior of local correlation
functions unveils universal features of the quantum dynamics ~\cite{CalabreseCardy2007,DS2011,
  SD2011, BD, BDV} and paves the way to the construction of effective field theories capable of
capturing them. Analytic results are relevant as benchmarks for cold atom experiments that very
recently~\cite{Esslinger, Esslinger_2, Esslinger_3} started to investigate particle and energy
transport under a unitary dynamics.

In this work, we study a quench which, although injecting an extensive amount of energy into the system, has mostly local effects.
 We take two uniformly filled long tight binding chains in
 their respective ground-states, but with a different number of fermions. At time $t=0$ the two edges are connected so that it turns into
 a single chain 
 \begin{equation}
\label{Ham}
H=-\frac{1}{2}\sum_{j=-L/2+1}^{L/2-1}\left(c^{\dagger}_j c_{j+1}+c^{\dagger}_{j+1}c_j \right).
\end{equation}
The hopping and interactions between sites $j=0$ and $j=1$ are initially absent;
the initial state $\ket{\psi_0}=\ket{\psi_l}\ket{\psi_r}$ is the tensor product of the ground states
(with specified occupancies) of the two decoupled chains. 
For unequal fillings, one expects some particle current at time $t>0$.  We denote by $k_F^l$
($k_F^r$) the Fermi momenta on the left (right), so that the particle number is $\frac{k_F^l
}{\pi}\frac{L}{2}$ ($\frac{k_F^r}{\pi}\frac{L}{2}$) on the left (right).  For simplicity we focus on
the symmetric case $k_F^l+k_F^r=\pi$, but extension to other values is straightforward.
It is useful to keep two limits in mind. When the fillings are equal $k_F^l=k_F^r=\pi/2$ there is no
particle current.  This particular quench was studied in
\cite{EKPP2008,CalabreseCardy2007,DS2011,SD2011}, using low-energy field theory, and belongs to the
class of Fermi-edge problems~\cite{Anderson,Affleck}.  The other simple limit is $k_F^l=\pi$, with a
domain wall (DW) initial
state~\cite{Antal_Schuetz_PRE1999,Antal_Krapivsky_Rakos_PRE2008,EislerRacz_PRL2013}
$\ket{\psi_0}=\prod_{x\leq 0} c_x^\dag\ket{0}$, where $\ket{0}$ is the fermion
vacuum. 
For intermediate filling a non equilibrium steady state (NESS) develops in the middle of the chain,
in which correlations are that of a Fermi sea $k\in [-k_F^r,k_F^l]$ with shifted momenta. This
central region may also be understood using bosonization~\cite{Lancaster_Mitra_PRE2010}.

The motivation for the present study is two-fold. First, our initial state is more refined than previously studied ``projected Fermi seas''\cite{Antal_Schuetz_PRE1999,Antal_Krapivsky_Rakos_PRE2008, Misguich_PRB2013}, and fully accounts for  the boundary effects near the junction. Second, we derive a 
number of exact results after the quench, in many interesting physical regimes.  These include the
long-time limit of correlations at finite $x$, the boosted Fermi sea regime, where we are also able
to derive the leading corrections. One of our main results concerns the regime at large positions
$x,y$ and time $t$, with finite ratios $x/t,y/t$. It is known that such regimes are qualitatively
well described by semiclassical arguments.  One complication in our case is that we are dealing with
a discrete lattice model with non polynomial dispersion, where usual techniques based on the use of
Wigner functions (see e.g \cite{BW11, BettelheimGlazman, Prot13}) are not easy to generalize. We
show nevertheless, by a careful use of stationary phase arguments in presence of singularities, that
such a picture becomes \emph{exact} in the thermodynamic limit, for $|x-y|\ll t$.  The present work
provides a simple unified picture of all physical regimes, in terms of initial state correlations
and single particle energies.  We also consider the Loschmidt echo, and combine analytical and
numerical techniques to treat the short- and large-time behaviors.  The Loschmidt echo is also
considered in the presence of nearest-neighbor interactions.

\paragraph*{Power series expansion---} 
We start by deriving an exact power series representation for the two point function.  All
higher order correlations may be obtained from Wick's theorem.
Consider two fermionic quadratic forms 
$A=\sum_{i,j=1}^{L} \mathcal{A}_{ij}c^{\dagger}_ic_j$ and $H=\sum_{i,j=1}^L \mathcal{H}_{ij}c^{\dagger}_ic_j$, where
$\mathcal{A}$ and 
$\mathcal{H}$ are $L\times L$ matrices. As is well known,
see e.g.~\cite{Klich}, quadratic forms in  Fock spaces are
a representation of the Lie algebra $gl(L)$, namely
$[A,H]=\sum_{i,j=1}^L[\mathcal{A},\mathcal{H}]_{i,j}c^{\dagger}_ic_j$.  Interpreting $H$ as the
final Hamiltonian and $A$ as an observable, application of the Baker-Campbell-Hausdorff formula
gives
\begin{equation}
\label{power_series}
 \langle\psi_0|A(t)|\psi_0\rangle=\sum_{N=0}^{\infty}\frac{(-it)^N}{N!}\tr\bigl[C_0^T~\ad_{\mathcal{H}}^{N}(\mathcal{A})\bigr],
\end{equation}
where
$\ad_{\mathcal{H}}^{N}(\mathcal{A})=[\dots[\mathcal{A},\overbrace{\mathcal{\mathcal{H}}],\dots,]\mathcal{H}}^{\text{N
    times}}]$ is a nested commutator, and $C_0^T$ is the transpose of the correlation matrix in the
initial state $[C_0]_{ij}=\langle\psi_0|c^{\dagger}_ic_j|\psi_0\rangle$.  If the matrix $C_0$ is
known, the representation in power series \eqref{power_series} can be efficiently computed.  Due to
the locality of $A$ and $H$, only $O(N^2)$ matrix elements in the nested commutator are
non-vanishing.  One can thus use a finite block of the correlation matrix $C_0$.  The coefficients
will then be exact in the thermodynamic limit as long as the elements in the nested commutator do
not reach the boundary of the block.  The method can be applied to derive exact expressions for any
two-point function, and also for the Loschmidt echo.  The correlation matrix in the initial state
$\ket{\psi_0}$ is $[C_0]_{ij}=C^r_{ij}$ for $i,j>0$ and $[C_0]_{ij}=C^l_{1-i,1-j}$ for $i,j\leq 0$,
with matrix elements $C^{\alpha}_{ij}$ ($\alpha=l,r$) given by
\begin{equation}
\label{cor_zero}
 C^{\alpha}_{ij}=S^{\alpha}(i-j)+S^{\alpha}(i+j),
\end{equation}
and $S^{\alpha}(x)=\sin{(k_F^{\alpha}x)}/(\pi x)$. The matrix $C^{\alpha}_{ij}$ is
a sum of a Toeplitz matrix, with entries that depend on  $i-j$, and a Hankel matrix, with entries that depend
on $i+j$.

\paragraph*{Long times, semiclassical regime and stationary phase approximation --}
\begin{figure}[htbp]
 \centering
 \includegraphics[width=0.97\columnwidth]{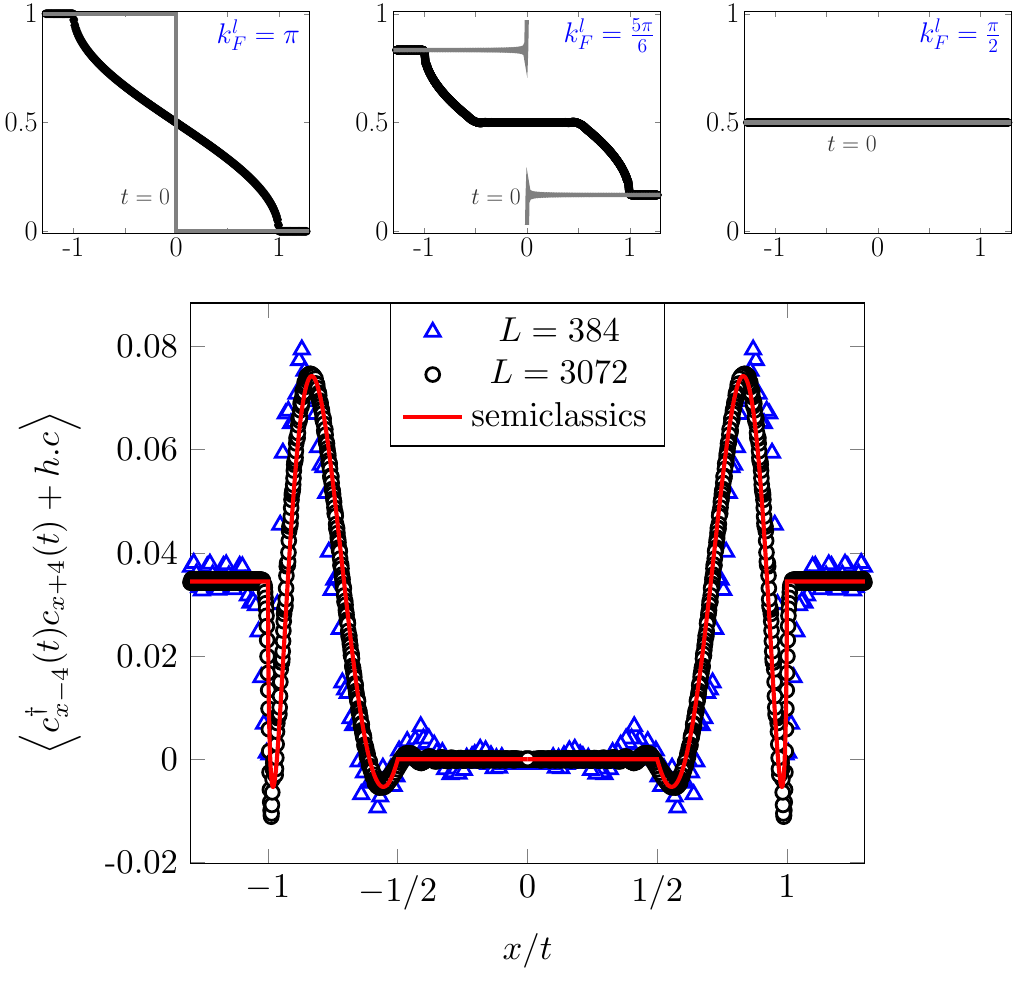}
 \caption{\emph{Top.} Particle density as a function of $x/t$ for $t=384$ and $L=3072$.
 The filling fractions are $k_F^l=\pi$, $k_F^l=5\pi/6$, $k_F^l=\pi/2$ from left to right. \emph{Bottom.}
 Comparison between semiclassics, Eq.~(\ref{eq:semiclassicsgeneral}),
 and finite-size numerical simulations for a correlator $\braket{c_{x-4}^\dag(t)c_{x+4}(t)}$. The fillings are $k_F^l=5\pi/6$, $k_F^r=\pi/6$. There are three regions. The first $|x/t|<1/2$ is the NESS. The correlations in the second region $|x/t|>1$ are that of the initial state. Most interesting is the third region $1/2<|x/t|<1$, with position dependent local correlations.}
 \label{fig:semiclassics}
\end{figure}
In the limit $L\to \infty$ the final Hamiltonian $H=\int dk~\varepsilon(k) f^\dag(k) f(k)$ is diagonal in momentum space, so the two-point function at time $t$ is
\begin{equation}
\label{Fourier}
 \Braket{c_x^\dag(t) c_{y}(t)}=\int \frac{dk \,dq}{2\pi}\, e^{i (\varepsilon(k)-\varepsilon(q))t
 -ikx+i q y} f(k,q),
\end{equation}
where $f(k,q)=\braket{\psi_0|f^{\dag}(k)f(q)|\psi_0}$ and $\varepsilon(k)=-\cos k$.  The integral is
taken over $[-\pi;\pi]^2$.  The large-time behavior for fixed $x$ and $y$ is then
determined~\cite{Barthel} from the points where the phase is stationary, as well as possible
singularities in $f(k,q)$. For this type of protocol, it is for example known that a NESS develops
in the middle~\cite{Antal_Krapivsky_Rakos_PRE2008,Lancaster_Mitra_PRE2010,Misguich_PRB2013}.  Here
we are interested in a different regime at large time where $x/t$ is kept finite. Such a limit has
been studied in several particular cases (including the DW
limit~\cite{Antal_Schuetz_PRE1999,Antal_Krapivsky_Rakos_PRE2008} or Fermi gases in the
continuum~\cite{BW11, BettelheimGlazman, Prot13}), but we present a general treatment for all
fillings here.  First we compute exactly $f(k,q)$, the Fourier transform of the initial correlation
matrix \eqref{cor_zero}.  We find \cite{suppl} that it is given by $f(k,q)=f^l(k,q)+f^r(k,q)$, where
each $f^{l/r}(k,q)=f^{l/r}_{T}(k,q)+f^{l/r}_{H}(k,q)$. The subscripts $T$ and $H$ refer to the
Toeplitz+Hankel decomposition of the matrix $C_{ij}^{l/r}$. These are given by
\begin{align}
\label{FT}
&f^l_{T}(k,q)=\frac{\chi_l(q)\;+\;g_l(e^{ik})-g_l(e^{iq})}{2\pi(1-e^{i(q-k+i0^+)})},\\
\label{FH}
&f_{H}^l(k,q)=\frac{e^{-iq}g_l(e^{-iq})-
 e^{ik}g_l(e^{ik})}{2\pi (e^{-ik}-e^{iq})},
\end{align}
where  $g_{l}(z)=\frac{i}{2\pi}\log\left[\frac{e^{ik^l_F}-z}{e^{-ik^l_F}-z}\right]$.
Here $\chi_l(k)=1$ if $k\in[-k^l_F,k_F^l]$ and zero otherwise; similar expressions hold for $f^{r}(k,q)$. Note that the $f^{l/r}_T$ have a pole and are not analytic at $\pm k_F^{l/r}$.
With this at hand, the asymptotic behavior of \eqref{Fourier} may be determined by using the stationary phase method.
The stationary points of the phase in \eqref{Fourier} satisfy the equations $v(k_s)-x/t=0$ and $v(q_s)-y/t=0$ for general $x$ and $y$, where $v(k)=\frac{d\varepsilon(k)}{dk}$
 is the group velocity. In the limit $x/t,y/t$ finite and $|x-y|/t\ll 1$ the two stationary points almost coincide and the integrand $f(k,q)$ in (\ref{FT}, \ref{FH}), is singular. The integral is then dominated by the region where 
 $k-q$ is small. Introducing new variables $K=(k+q)/2$ and $Q=k-q$, the stationary point is located at $Q_s=0$,
 irrespective of $K$. Expanding around it and using $\int_\mathbb{R} \frac{dQ}{2\pi i} \frac{e^{i Q x}}{Q-i0^+}=\Theta(x)$ we obtain
 \begin{align}\nonumber
  \langle c_x^\dag(t)&c_y(t)\rangle=\int_{-k_F^l}^{k_F^l} \frac{dK}{2\pi} e^{-i K (x-y)}\Theta\left(-\frac{x+y}{2}+v(K)t\right)
  \\\label{eq:semiclassicsgeneral}
  &+\int_{-k_F^r}^{k_F^r} \frac{dK}{2\pi} e^{-i K (x-y)}\Theta\left(\frac{x+y}{2}-v(K)t\right),
 \end{align}
 where $\Theta(x)$ is the Heaviside step function. The density
 $\rho(x,t)=\braket{c_x^\dag(t)c_x(t)}$ \cite{Antal_Schuetz_PRE1999} is obtained by setting
 $x=y$. See Fig.~\ref{fig:semiclassics}, top panel; a numerical check at distance $|x-y|=8$ is also
 shown in the bottom panel. As can be seen the agreement is excellent and improves when increasing
 the system size, which confirms that it becomes exact in the limit $L\to \infty$. Note that
 (\ref{eq:semiclassicsgeneral}) is entirely determined by the pole contribution in $f(k,q)$, but
 subleading corrections depend on its full analytic form.

Such types of results are known as semiclassical (or hydrodymamic) approximations in the literature, as each fermion at momentum $k$ in the initial state propagates ballistically at speed $v(k)$. 
Eq.~(\ref{eq:semiclassicsgeneral}) is then the statement that this approximation becomes 
exact in the scaling limit with $x/t,y/t$ finite but $|x-y|/t\ll 1$. There are three regions.
In the first ($|x/t|>1$), one of the step function is identically zero while the other is identically one.
The correlations are locally those in the bulk of the initial left (or right) ground state.
In the central region $|x/t|<\sin k_F^l$ the correlations are those of a shifted Fermi sea with momenta
in $[-k_F^r,k_F^l]$. This steady state carries some current. Most interesting are the intermediate regions $\sin k_F^l<|x/t|<1$,
which are inhomogeneous. In that case the correlations are more complicated, as is shown in Fig.~\ref{fig:semiclassics}.
Another interesting feature is that a naive low energy field theory description would break down,
due to the inhomogeneous background. Indeed, the Fermi velocity depends on position and time, so it has to be a field theory in a curved space-time with a metric $ds^2= dx^2-v(x,t)^2 dt^2$. Since in 1+1d all metrics are conformally equivalent, it is always possible to come back to a flat geometry. However, the relation between lattice and continuous geometries is complicated. 
Such aspects are investigated in \cite{Arctic}. 
Semiclassical results can also be obtained for other physically relevant initial states, \textit{e.g.}
when the two halves are held at different temperatures $T_l$ and $T_r$.
The large $x$ and $t$ behavior is still dominated by the pole in $f(k,q)$~\cite{DMV14} with residue proportional
to the Fermi-Dirac distributions at temperatures $T_{l/r}$. The inhomogeneous energy density profile can be similarly computed \cite{suppl}. A numerical verification of the validity of
a semiclassical picture for energy transport in the continuum is presented in~\cite{CK14}.

Let us now discuss other large-time regimes. First, it is interesting to look at what happens close
to the edge of the front $x/t\simeq \pm 1$. For the DW limit it has been shown
\cite{EislerRacz_PRL2013} that suitably rescaled correlations are described by the Airy
kernel~\cite{TW93}.  Such a result follows readily from our formalism for \emph{any} filling
$k_F^l\neq \pi/2$ \cite{suppl}.  For large-distance correlations with $|x-y|/t$ of order one,
(\ref{eq:semiclassicsgeneral}) breaks down. The reason is that the two points $k_s$ and $q_s$ where
the phase is stationary do not coincide anymore. Our method can be generalized to such a case,
however, by expanding inside the exponentials around this point, and by careful treatment of the
boundary contributions at $\pm k_F^{l/r}$. In this case the computations become more involved, and
depend on the exact form of $f(k,q)$. Another example with boundary contributions is discussed
below.

\paragraph*{Corrections to the steady state --}
\begin{figure}[t]
\centering
\includegraphics[width=\columnwidth]{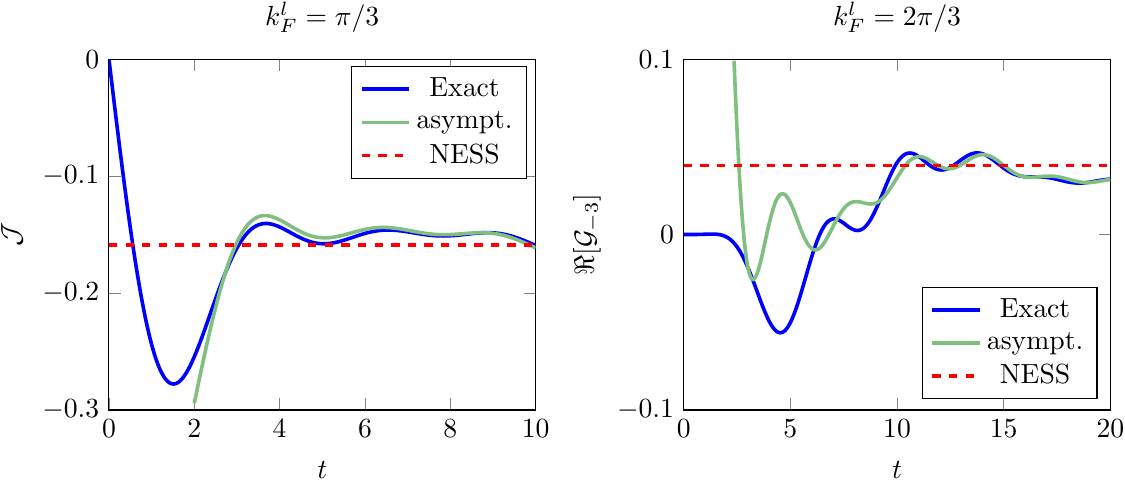}
\caption{\textit{Left.} Exact expression for the particle current (in blue) at fillings $k_F^l=\pi-k_F^r=\pi/3$
obtained from \eqref{current}. The green curve is the asymptotics expansion \eqref{asymptotic_J} derived with a stationary
phase approximation. Finally  the dashed red line is the value attained in the NESS by
the current. \textit{Right.}  Similar plot for the
real part of $\mathcal{G}_{-3}(t)\equiv\langle c^{\dagger}_{-3}(t)c_4(t)\rangle$ at $k_F^l=2\pi/3$.}
\label{fig:current}
\end{figure}
We now look at the corrections to the steady state predicted by Eq.~(\ref{eq:semiclassicsgeneral}), in the regime $x/t\ll 1$. The simplest example is the current $\mathcal{J}(t)={\rm Im} \braket{c_0^\dag(t) c_1(t)}$ through the link in the middle of the junction. From the power series method we obtain
\begin{multline}
\label{current}
\mathcal{J}(t)=2\sum_{k=1}^{\infty}\frac{(-1)^k}{(2k-1)!}\left(\frac{t}{2}\right)^{2k-1}\sum_{s=1}^{k}\bigg[
S^l(2s)\frac{(C_{2k-1}^{k+s-1})^2}{k+s}\\\times \frac{4ks-k-s}{2k-1}\bigg]
\;+\;\left(\frac{k_F^l}{\pi}-\frac{1}{2}\right)t [J_0^2(t)+J_1^2(t)],
\end{multline}
with $C_n^k$ the binomial coefficient $\binom{n}{k}$ and $J_\nu(t)$ the Bessel function of the first kind.
Eq.~\eqref{current} extends to general $k_F^l$ the result for the particle current
derived in~\cite{Antal_Krapivsky_Rakos_PRE2008}  for a DW initial state ($k_F^l=\pi$, where only the last term remains). 

The stationary phase treatment of this correlation requires some special care~\cite{Wong}, due to
the singularities in $f(k,q)$ and the presence of sharp boundaries in $k$-space when $k_F^l\neq
\pi$. The singularity at $k=q$ may be removed by studying the time-derivative of
\eqref{Fourier}. The extra factor $\varepsilon(k)-\varepsilon(q)$ gets rid of the denominator in
(\ref{FT}, \ref{FH}) and the integrations over $k$ and $q$ are now decoupled \cite{suppl}. The
semiclassical results may also be recovered using this derivative procedure. The resulting one-dimensional
integrals can be evaluated by stationary phase in the large $t$ limit and the result integrated back
to derive the asymptotic expansion of the two-point function. For the current through the junction
we find
\begin{multline}
\label{asymptotic_J}
\mathcal{J}(t) ~\stackrel{\small t\gg 1}{\sim}~ \frac{\cos(\pi-k_F^l)}{\pi}+
\frac{(\pi -2 k_F^l) \cos 2 t}{2 \pi ^2 t}-\\
\frac{ \sin(t_-) \sin (t \cos k_F)+\cos (k_F^l-t \cos k_F^l)\sin(t_+)}{2^{-1/2}\sin(k_F^l)(\pi
  t)^{3/2}},\end{multline} with $t_{\pm}=t\pm\pi/4$.  The interpretation goes as follows. The
approach to the NESS of arbitrary correlation functions contains semi-integer powers of $t$ for all
$k_F^l\not=0,\pi$; these fractional powers are due to the aforementioned boundary contributions, and
are universal.  Also, the frequency of oscillations are
$\omega(k_s,k'_s)=|\varepsilon(k_s)-\varepsilon(k_s')|$, where $k_s$ and $k'_s$ belong to the set of
critical points $\{0,\pi,k_F^l\}$, obtained from the stationary phase. A comparison between the
exact \eqref{current} and asymptotic \eqref{asymptotic_J} results is shown in
Fig.~\ref{fig:current}.

\begin{figure}[t]
\centering
\includegraphics[width=\columnwidth]{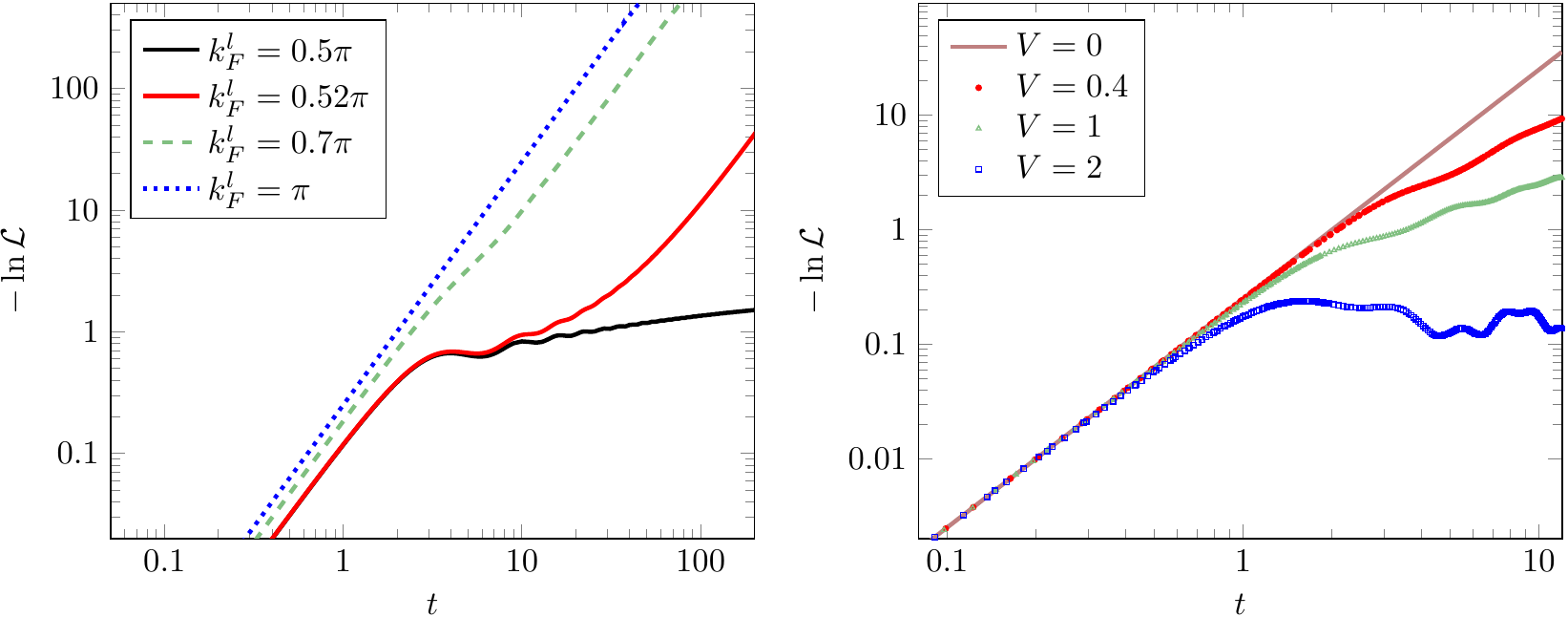}
\caption{\emph{Left}. $-\ln \mathcal{L}(t)$ for several fillings at $V=0$. \emph{Right}. Same with varying interaction strengths $V$ for
the DW initial state. The length of the system is $L=24$. Notice how in the gapped
phase the Loschmidt echo relaxes to a finite value and keeps instead decreasing with time in the gapless phase.}
\label{fig_lecho}
\end{figure}

\paragraph*{Loschmidt echo  --} 

The Loschmidt echo is defined as the overlap
$\mathcal{L}(t)=\left|\braket{\psi_0|e^{-iHt}|\psi_0}\right|^2$.  For short times, a power-series
expansion gives $\mathcal{L}(t)\approx1-\tilde{\gamma}{t^2}$, with
$\tilde{\gamma}=\langle{H^2}\rangle-\langle{H}\rangle^2$ the variance of the post-quench Hamiltonian
in the initial state; $\tilde{\gamma}$ can be computed exactly \cite{suppl}.

The long-time behavior was studied previously for $k_F^l=\pi/2$ \cite{DS2011,SD2011}, where only
low-energy excitations are generated, and conformal field theory techniques (CFT) may be used. At
times much larger than the lattice spacing, it decays as a universal power-law $\mathcal{L}(t)\sim t^{-c/4}$,
 a manifestation of the
celebrated Anderson orthogonality catastrophe~\cite{Anderson, Affleck}.
 $c$ is the central charge of the CFT (here $c=1$). In the DW limit, we
are able to compute $\mathcal{L}(t)$ exactly as follows. By application of the Wick theorem,
$\mathcal{L}(t)=\displaystyle{\lim_{n\to \infty}\det_{1\leq i,j\leq n}}\left(g_{i-j}-g_{i+j}\right)$, where
$g_p=\frac{1}{2\pi}\int_{-\pi}^{\pi}d\theta e^{-ip\theta+it \varepsilon(\theta)}$. 
Using a generalization~\cite{Basor} of the strong Szeg\H{o} limit theorem \cite{Szego} to 
Toeplitz+Hankel determinants, we are able to derive
\begin{equation}\label{eq:loschmidtexact}
 \mathcal{L}(t)=e^{-t^2/4} . 
\end{equation}
This can also be derived using the short-time expansion \cite{suppl}. 
This simple formula has several remarkable features. First, it is exact at \emph{all times} in an infinite system (for finite $L$, it is exponentially accurate until $t\simeq L/2$).
Another is that (the logarithm of) its imaginary time version $\mathcal{L}(i\tau)$, $\tau \in \mathbb{R}$, can be
interpreted~\cite{Arctic} as the free energy of
a statistical mechanical system where all degrees of freedom outside of
a circle of radius $\tau/2$ are frozen, analogous
to the celebrated ``arctic circle theorem''~\cite{JockushProppShor} for classical dimers.

At intermediate fillings, the long-time behavior can only be accessed numerically. 
We find a gaussian decay for any $k_l>\pi/2$, i.e., $\mathcal{L}(t)\sim e^{-\gamma{t^2}}$, with a prefactor well described by the formula
$\gamma=\frac{1}{4}\cos^2k_F^l$. 
The short- and long-time behaviors of $\mathcal{L}(t)$ are summarized in Fig. \ref{fig_lecho}
for several fillings, through plots of $-\ln \mathcal{L}(t)$ versus time.

\paragraph*{Finite interactions --} 
Let us add to the Hamiltonian \eqref{Ham} the density-density interaction $V\sum_{j}n_{j}n_{j+1}$
with $V>0$ and focus on the DW initial state.  In the gapless phase ($0\leq V\leq 1$), the
stationary state supporting a ballistic particle current has been numerically investigated in
\cite{KollathSchuetz_PRE2005, Misguich_PRB2013, Alba_HeidrichMeisner_PRB2014}.  Correlation
functions in the NESS are those of the ground-state at half-filling multiplied by a space-dependent
phase~\cite{Lancaster_Mitra_PRE2010}.  In the gapped phase, the presence of heavy-mass Hamiltonian
eigenstates having dominant overlap with the DW initial state~\cite{MC_DW} prevents the formation of
a light-cone and leads eventually to absence of particle transport for large
times~\cite{KollathSchuetz_PRE2005}.
All these features are also transparent in the Loschmidt echo, which we plot in
Fig. \ref{fig_lecho}. To avoid recurrences we assume $t<L/2$ and $L$ sufficiently large. In the
gapless phase, classical free-energy arguments~\cite{KZ00} outlined above suggest
the large-time behavior $-\ln \mathcal{L}(t)\stackrel{t\gg 1}{\sim} a(V)t^2$, where $a(V)$ is a coefficient. In
the gapped phase $\mathcal{L}(t)$ relaxes at large times to a finite $L$-dependent value. Such a
value is expected to vanish as $L\rightarrow\infty$.  Relaxation features an oscillatory behavior as
a consequence again of the presence of slow bound states with large scalar product with our initial
state.  

For small times, the Loschmidt echo behaves as $1-t^2/4$ for all $V$ \cite{MC_DW, TS14}, since the
only process contributing to the cumulant $\langle H^2\rangle-\langle H\rangle^2$ corresponds to
moving the rightmost particle one step to the right, and then back.

\paragraph*{Conclusion--}
In this Letter we presented a unified formalism to characterize both the short- and long- time limit of correlation
functions in a  fermionic chain. Our main example was a system with a inhomogeneous density profile but the method applies
to a wide class of initial states.
Together with exact expressions for the particle current and the Loschmidt echo,
we provided a full analytical derivation of the so-called semiclassical regime and outlined its domain of validity.
We showed how subleading corrections to correlations at the edges of the front are described  by the Airy kernel, giving 
support to a connection between  large deviation functions for transport problems
in fermionic systems and the Tracy-Widom distribution.
Our work raises  a number of interesting questions: among them, the possibility of a field-theory formulation of the scaling behavior,
the study of the entanglement spreading in
a inhomogeneous background and a deeper understanding of interaction effects, which might be achieved by
engineering a tailored Bethe rapidity distribution for the NESS.

\paragraph*{Acknowledgments.}

We wish to thank M.~Brockmann, P.~Calabrese, P.~Krapivsky, C.~Krattenthaler, G.~Misguich, A.~Mitra, L.~Santos and
G.~Sch\"utz for stimulating discussions. JV and JMS acknowledge hospitality and support from the
Galileo Galilei Institute during the program "Statistical Mechanics, Integrability and
Combinatorics" in Florence. JD thanks the Max Planck Visitors Program for hospitality and support during his stay at MPIPKS.

\pagebreak


\onecolumngrid
\setcounter{equation}{0}%
\setcounter{figure}{0}%
\setcounter{table}{0}%
\renewcommand{\thetable}{S\arabic{table}}
\renewcommand{\theequation}{S\arabic{equation}}
\renewcommand{\thefigure}{S\arabic{figure}}

\newcounter{app}
\setcounter{app}{0}
\newcounter{app2}
\setcounter{app2}{1}
\newcommand{\appsection}[1]{\setcounter{app2}{1}\stepcounter{app}\section{\Alph{app}.\hspace{1ex}#1}}
\newcommand{\appsubsection}[1]{\subsection{\Alph{app}\arabic{app2}.\hspace{1ex} #1}\stepcounter{app2}}

\begin{center}
{\Large Supplementary Material for EPAPS \\ 
\titleinfo
}
\end{center}
This supplementary material contains additional information about
\begin{itemize}
 \item The calculation of the function $f(k,q)$ in the inhomogeneous density quench (Sec. \hyperlink{sec:fkq}{A})
 \item The power series expansion of  correlation functions (Sec. \hyperlink{sec:power}{B})
 \item The stationary phase approximation of correlation functions (Sec. \hyperlink{sec:statphase}{C})
 \item The Loschmidt echo (Sec. \hyperlink{sec:Loschmidt}{D})
\end{itemize}

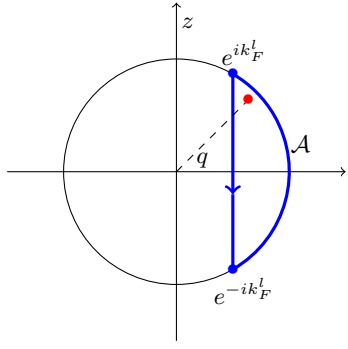
\begin{figure}[t]
\begin{tikzpicture}[scale=1.5]
 \draw[->] (0,-1.5)--(0,1.5);
 \draw[->] (-1.5,0)--(1.5,0);
 \draw (0,0) circle [radius=1];
 \draw[blue, very thick] (0.5, -1.732*0.5) arc [radius=1, start angle= 60-90-30, end angle=150-90];
 \draw[blue, very thick, ->] (0.5, 1.732*0.5)--(0.5, -0.2);
 \draw[blue, very thick] (0.5, -0.2)--(0.5,- 1.732*0.5);
 \draw[dashed] (0,0)--(0.9*1.414/2, 0.9*1.414/2);
 \draw(0.9*1.414/2, 0.9*1.414/2) node[red] {$\bullet$};
 \draw(0.1,0.1) node[right] {$q$};
 \draw(0.1, 1.2) node [above] {$z$};
 \draw(0.5, -1.732*0.5) node[blue]{$\bullet$};
 \draw(0.5, 1.732*0.5) node[blue]{$\bullet$};
 \draw(0.6, -1.732*0.5) node[below]{$e^{-ik_F^l}$};
 \draw(0.6, 1.732*0.5) node[above]{$e^{ik_F^l}$};
 \draw(1.1,0.1) node[above] {$\mathcal{A}$};
 \end{tikzpicture}
\caption{Integration contour $\mathcal{A}$ utilized to evaluate the integrals in (\ref{SfT}, \ref{SfH}).
Consider for example
the function $f^l_T(k,q)$ in \eqref{SflT}, then the only singularity inside the contour $\mathcal{A}$ is at $p=q+i0$. }
\label{fig_contour}
\end{figure}

\hypertarget{sec:fkq}{\appsection{Calculation of the function $f(k,q)$ in the inhomogeneous density quench}}
Before connecting them the two fermionic Hamiltonians have open boundary conditions. In the thermodynamic limit
$L\rightarrow\infty$, they are separately diagonalized in momentum space by the following transformations
\begin{align}
 &c_j=\sqrt{\frac{2}{\pi}}\int_{0}^{\pi}dp~\sin(pj)f_r(p)\quad j>0,\\
 &c_j=\sqrt{\frac{2}{\pi}}\int_{0}^{\pi}dp~\sin(p(j-1))f_l(p)\quad j\leq 0,
\end{align}
where the fermionic operators $f^{\dagger}_{r/l}(p)$ create a fermion with momentum $p$ on the right and left half of the chain
respectively. The one particles states \textit{before} the quench
are therefore $|p\rangle_{r/l}\equiv f^{\dagger}_{r/l}(p)|0\rangle$. \textit{After} the quench, the Hamiltonian is fully translation
invariant and diagonalized by Fourier transform, i.e. the local fermions $c_j$ are expressed by
$c_j=\frac{1}{\sqrt{2\pi}}\int_{-\pi}^{\pi}dk~e^{ikj}f(k)$ and the one-particle state \textit{after} the quench are given by
$|k\rangle\equiv f^{\dagger}(k)|0\rangle$. The matrix elements between the single-particle states before and after the quench are 
\begin{align}
\label{SMr}
&M_{r}(k,p)\equiv~\langle k|p\rangle_{r}=\frac{1}{2\pi i}\left[\frac{1}{1-e^{i(k-p-i0)}}-(p\rightarrow -p)\right],\\
\label{SMl}
&M_{l}(k,p)\equiv~\langle k|p\rangle_{l}=\frac{1}{2\pi i}\left[\frac{e^{-ip}}{1-e^{i(k-p+i0)}}-(p\rightarrow -p)\right].
\end{align}
The initial state $|\psi_0\rangle$ we chose is the factorized Fermi sea $|\psi_l\rangle|\psi_r\rangle$ and one simply has
$\langle\psi_0|f^{\dagger}_{\alpha}(p)f_{\beta}(p')|\psi_0\rangle=\delta(p-p')\delta_{\alpha,\beta}\Theta(k_F^{\alpha}-p)$,
for $\alpha,\beta=\{r,l\}$. Then $f(k,q)$ in \eqref{Fourier} is the sum $f^l(k,q)+f^r(k,q)$ where 
\begin{equation}
\label{Sf}
 f^{\alpha}(k,q)=\int_{0}^{k_F^{\alpha}}dp~M_{\alpha}^*(k,p)M_{\alpha}(q,p).
\end{equation}
The integration domain in \eqref{Sf} can be extended to $p\in[-k_F^{\alpha},k_F^{\alpha}]$, exploiting the symmetry
properties of the functions in (\ref{SMr},~\ref{SMl}). It is also useful to further decompose $f_{\alpha}(k,q)$ into the sum
$f^{\alpha}_T(k,q)+f^{\alpha}_H(k,q)$. For example one has
\begin{align}
\label{SfT}
f^l_T(k,q)= \frac{1}{4\pi^2}\int_{-k_F^l}^{k_F^l}dp~\frac{1}{(1-e^{-i(k-p-i0)})(1-e^{i(q-p+i0)})},\\
\label{SfH}
f^l_H(k,q)=-\frac{1}{4\pi^2}\int_{-k_F^l}^{k_F^l}dp~\frac{e^{2ip}}
 {(1-e^{-i(k-p-i0)})(1-e^{i(q+p+i0)})},
\end{align}
and very similar expressions for  $f^r(k,q)$. Notice that the subscripts $T$ or $H$ refer to having chosen products of
terms in (\ref{SMr},~\ref{SMl}), singular for $k=\pm p$ and $q=\pm p$ (subscript $T$) or $k=\pm p$ and $q=\mp p$
(subscript $H$). The integrals in (\ref{SfT},\ref{SfH}) can be computed changing variable $z=e^{ip}$ and performing
integration along the contour $\mathcal{A}$ depicted in Fig.~\ref{fig_contour}. Only the singularities inside the contour contribute to the final result.

Performing those integrations we obtain
\begin{align}
\label{SflT}
 &4\pi^2 f^l_T(k,q)=\frac{i}{1-e^{i(q-k+i0)}}
 \left[\log\left(\frac{e^{ik_F^l}-e^{ik}}{e^{-ik_F^l}-e^{ik}}\right)-
 \log\left(\frac{e^{ik_F^l}-e^{iq}}{e^{-ik_F^l}-e^{iq}}\right)-2\pi i\chi_l(q)\right],\\
\label{SflH}
 & 4\pi^2 f^l_H(k,q)=\frac{i}{e^{-ik}-e^{iq}}
 \left[e^{-iq}\log\left(\frac{e^{ik_F^l}-e^{-iq}}{e^{-ik_F^l}-e^{-iq}}\right)-
 e^{ik}\log\left(\frac{e^{ik_F^l}-e^{ik}}{e^{-ik_F^l}-e^{ik}}\right)\right],\\
 \label{SfrT}
 &4\pi^2 f^{r}_T(k,q)=\frac{i}{1-e^{i(q-k-i0)}}\left[\log\left(\frac{e^{ik_F^r}-e^{ik}}{e^{-ik_F^r}-e^{ik}}\right)-
 \log\left(\frac{e^{ik_F^r}-e^{iq}}{e^{-ik_F^r}-e^{iq}}\right)+2\pi i\chi_r(k)\right]\\
 \label{SfrH}
 &4\pi^2 f^{r}_H(k,q)=\frac{i}{e^{-ik}-e^{iq}}\left[e^{-ik}\log\left(\frac{e^{ik_F^r}-e^{-ik}}{e^{-ik_F^r}-e^{-ik}}\right)-
 e^{iq}\log\left(\frac{e^{ik_F^r}-e^{iq}}{e^{-ik_F^r}-e^{iq}}\right)\right];
\end{align}
where $\chi_{\alpha}(x)$, is a function with value one if $x\in[-k_F^{\alpha},k_F^{\alpha}]$ and zero otherwise. The initial state
correlation matrix $[C_0]_{nm}=\langle\psi_0|c_{n}^{\dagger}c_m|\psi_0\rangle$ is given by
\begin{equation}
\label{Scorr_0}
[C_0]_{nm}=\overbrace{\Theta(n-1)\Theta(m-1)[S^r(n-m)+S^r(n+m)]}^{C^r_{nm}}+\overbrace{\Theta(-n)\Theta(-m)
[S^l(n-m)+S^l(-n-m+2)]}^{C^l_{1-n, 1-m}},
\end{equation}
where $S^{\alpha}(x)$ has been defined in the main text below \eqref{cor_zero}. It is possible to further check
(\ref{SflT}-\ref{SfrH}), showing that they yield indeed the Fourier transform of $[C_0]_{nm}$
\begin{equation}
\label{check_F}
[C_0]_{nm}=\frac{1}{2\pi}\int_{-\pi}^{\pi}dk \int_{-\pi}^{\pi}dq~ e^{-ikn+i q m} f(k,q), 
\end{equation}
as it follows from \eqref{Fourier}. To verify \eqref{check_F} it might be useful the
following identity valid for $s\in\mathbb Z$ and $a\in \mathbb R$
\begin{equation}
\int_{\mathcal{C} }dz~z^{s-1}\log\left(\frac{e^{ia}-z}{e^{-ia}-z}\right)=
4\pi\Theta(s-1)\frac{\sin(as)}{s},
\end{equation}
with $\mathcal{C}$ the unit circle.

\begin{figure}[t]
\centering
\includegraphics[width=6cm]{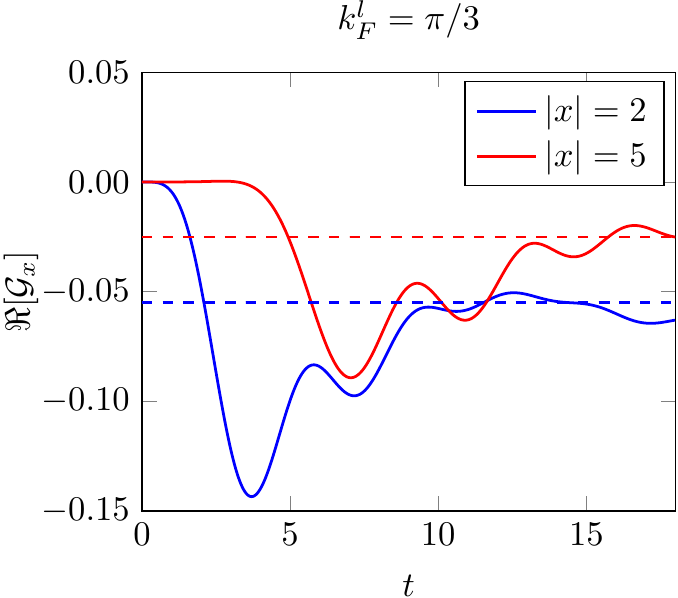}
\caption{Real part of the two-point functions $\mathcal{G}_{x}(t) = \langle c^{\dagger}_{-x}(t)c_{x+1}(t)\rangle$ plotted for different values of $|x|=2,5$ at
$k_F^l=\pi-k_F^r=\pi/3$. The curves
are obtained from the power series expansion \eqref{Scoeff} summing  $N=255$ coefficients. We observe clear signatures of
light-cone effects at short times and relaxation to the stationary value in \eqref{SNESS},
denoted by dashed lines in the figure, at larger times.}
\label{fig_cc}
\end{figure}

\hypertarget{sec:power}{\appsection{Power series expansion of correlation functions}}
An exact expression for the coefficients in the power series expansion of the correlation functions in
\eqref{power_series} can be obtained starting from the knowledge of the nested commutator $\ad_{\mathcal H}^N(\mathcal{A})$.
For the Hamiltonian \eqref{Ham} and $A=c^{\dagger}_{x}c_y$ simple combinatorics leads to  
\begin{equation}
\label{comm_exact}
 [\ad_{\mathcal{H}}^N(\mathcal{A})]_{ij}=\frac{(-1)^{i-x+N}}{2^N}\binom{N}{\frac{N+|d-(x-y)|}{2}}\binom{N}{\frac{N+|D-(x+y)|}{2}},
\end{equation}
where $d=i-j$ and $D=i+j$, satisfying the constraints $|d-(x-y)|\leq N,~|D-(x+y)|\leq N $
and  $N-(x-y)\equiv D~\text{mod}~2$. Moreover when $y=-x+1$ ($x\leq 0$) the points are taken symmetrically
on the left and right halves  and it is not difficult to prove that all the coefficients  in \eqref{power_series} are
vanishing if $N\leq |x|$, which is a consequence of light-cone effects.
From \eqref{power_series} it is clear that
\begin{equation}
\label{Spower}
 \langle c^{\dagger}_x (t)c_y(t)\rangle=\sum_{N=0}^{\infty}\frac{(-i)^N t^N}{N!}(a_r^N+a_l^N),
\end{equation}
where $a_{\alpha}^N=\Tr[C_\alpha^T\ad_{\mathcal H}^N(\mathcal{A})]$ and the matrices $C_{\alpha}$ defined in \eqref{Scorr_0} for
$\alpha=l,r$. Let us focus again on the symmetric case $x=-y+1$ and $|x|=-x$; 
then one has 
\begin{align}
\nonumber
 a_r^{N}&=\sum_{n,m\geq 1}C^r_{nm}[\ad_\mathcal{H}^N(\mathcal{A})]_{nm}=
 (-1)^{|x|}\sideset{}{'}\sum_{d=-N+1}^{N-2|x|-1}(-1)^{d/2}
 S^r(d)\binom{N}{\frac{N+|d+2|x|+1|}{2}}
 \sideset{}{'}\sum_{D=2+|d|}^{N+1}(-1)^{D/2}\binom{N}{\frac{N+D-1}{2}}\\
 \label{Scoeff}
 &+(-1)^{|x|}\sideset{}{'}\sum_{D=2+\text{mod}(N+1,2)}^{N+1}(-1)^{D/2}S^r(D)\binom{N}{\frac{N+D-1}{2}}
 \sideset{}{'}\sum_{d=-(D-2)}^{\text{min}(D-2,N-2|x|-1)}(-1)^{d/2}\binom{N}{\frac{N+|d+2|x|+1|}{2}}
\end{align}
with the primed sum restricted to  $d\equiv N+1\mod 2$ and $D\equiv N+1\mod 2$. The expression \eqref{Scoeff} can be simplified
in the case $x=0$. The coefficients $a_{r/l}^{2k-1}$ are the coefficients in the power series expansion of the particle
current $\mathcal{J}$, we get
\begin{align}
 a_r^{(2k-1)}=-\sideset{}{'}\sum_{s=0}^{k-1}
 \left[\frac{(k+s) S^r(2s)}{2k-1}C_{2k}^{k+s}C_{2k-1}^{k+s}+S^r(2s+2)(C_{2k-1}^{k+s})^2\right]
\end{align}
where the primed summation denotes now that the term containing $S_r(0)$ has to be divided by two. We also introduced the
notation $C_{n}^k$ for the binomial coefficient $\binom{n}{k}$.
Analogously,
one finds the odd coefficients $a_l^{(2k-1)}$ and  summing the two, see \eqref{Spower}, we derive the result
given in \eqref{current} for the particle current. Although \eqref{Scoeff} appears complicated the coefficients
of the power series expansion can be derived in symbolic form for all  two-point functions.
We show in Fig. \ref{fig_cc} typical plots of the symmetric correlators
$\mathcal{G}_{x}=\langle\psi_0|c^{\dagger}_{x}(t)c_{-x+1}(t)|\psi_0\rangle$
obtained with the method above.
The
signal starts to be non-vanishing for $t\sim |x|$ which is a clear indication of light-cone effects.
Correlation functions approach their stationary value at long times (dashed in curves in Fig. \ref{fig_cc})
\begin{equation}
\label{SNESS}
\langle c^{\dagger}_{x}c_{y}\rangle_{\rm NESS}=e^{-i\delta (x-y)}\frac{\sin (k_0 (x-y))}{\pi(x-y)} 
\end{equation}
for $\delta=k_F^l-k_0$ and $k_0=\pi/2$ that will be derived in Sec. C2.

\hypertarget{sec:statphase}{\appsection{Stationary phase approximation of correlation functions}}
The asymptotics of correlation functions $\langle c^{\dagger}_x(t) c_y(t)\rangle$
can be studied mainly in three interesting regimes determined by the value of the parameters $x,y$ and $t$
entering the Fourier transform \eqref{Fourier}. We will consider the cases
\begin{itemize}
 \item$x/t\rightarrow 0$ and $y/t\rightarrow 0$ with $t\gg 1$,
 see Sec.~\hyperlink{sec:corrections}{C1}.
 The points $x$ and $y$
 are inside the light-cone and correlations have already relaxed to
 their stationary value in the translation invariant NESS. The approach to the stationary state is oscillatory
 with corrections organized in  power series of $t^{-1/2}$.
\item $x/t$ and $y/t$ finite but all the variables large, see Sec.~\hyperlink{sec:semiclassics}{C2}. This particular limit is known in the literature as the
\textit{semiclassical (or scaling) limit}; we offer an alternative derivation of our main results in
Sec.~\hyperlink{sec:semiclassicsbis}{C3}.
\item $x/t\simeq \pm 1$, see Sec.~\hyperlink{sec:Airy}{C4}. We focus on the behavior of  correlations
at the edge of the fronts and show that they are described
by the Airy kernel. 
 \end{itemize}
\hypertarget{sec:corrections}{\appsubsection{Corrections to the steady state}}
We start considering the first case that technically amounts to performing a stationary phase approximation in
the variable $t$ of the double integral \eqref{Fourier}. The function $f(k,q)$ is singular for $k=q$ and we found it easier to
consider first the time derivative of the two-point correlation function. The extra factor $\varepsilon(k)-\varepsilon(q)$
exactly cancels the divergence in both $f^{l/r}_{T}(k,q)$ and $f^{l/r}_{H}(k,q)$ in (\ref{SflT}-\ref{SfrH}). Moreover the integrations over $k$ and $q$ are now
decoupled and one needs to analyze the large-time behavior of the following one-dimensional integrals
\begin{align}
\label{SA}
&A(n,t)=\frac{1}{2\pi}\int_{-\pi}^{\pi} dk\, e^{it\cos k}e^{ink},\\
\label{SB}
&B(k_F,n,t)=\frac{1}{2\pi}\int_{-k_F}^{k_F} dk\, e^{it\cos k}e^{ink},\\
\label{SC}
&C(k_F,n,t)=\frac{1}{2\pi}\int_{-\pi}^{\pi} dk\, e^{it\cos k}e^{ink} \log \left(\frac{e^{ik_F}-e^{ik}}{e^{-ik_F}-e^{ik}}\right),
\end{align}
which is a more comfortable situation. We will provide asymptotic expansion of correlators up to $O(t^{-3/2})$, that implies
that we have to keep terms in the asymptotics expansions of the derivative up to $O(t^{-5/2})$. Since the contribution
of ordinary stationary points is of the form $t^{-p/2}$ for $p$ positive integer, we have to expand the functions
(\ref{SA}-\ref{SC}) up to order $t^{-2}$
The first integral \eqref{SA}  is proportional to a Bessel function whose asymptotics are known. However it
is good exercise to recover it. There are two stationary points at $k=0$ and $k=\pi$ and the result is
\begin{equation}
\label{S_as_a}
 A(n,t)=\frac{e^{in\pi/2}\sqrt{2}\sin \left(\frac{\pi}{4}-\frac{n\pi}{2}+t\right)}{\sqrt{\pi t}}+\frac{4n^2-1}{8}\frac{e^{in\pi/2}\sqrt{2}\sin \left(\frac{\pi}{4}+\frac{n\pi}{2}-t\right)}{\sqrt{\pi} t^{3/2}}+O(t^{-5/2})
\end{equation}
Notice that the asymptotics expansion of $A(n,t)$ only contains seminteger powers of  $t$. To derive exact asymptotics for the
function $B(k_F,n,t)$ one has to add the boundary terms at $k=\pm k_F$ to
the contribution from the stationary point located at $k=0$.
Boundary contributions can be obtained by making the change of variables $u=\cos k$ and then
integrating by parts~\cite{Wong}. These boundary terms generate integer powers of $t$ in the expansion
\begin{equation}
\label{S_as_B}
 B(k_F,n,t)=\frac{e^{i(t-\pi/4)}}{\sqrt{2\pi t}}+\frac{i e^{i t \cos k_F} \cos nk_F}{\pi t \sin k_F}+
 \frac{4n^2-1}{8}\frac{e^{i(t+\pi/4)}}{\sqrt{2\pi} t^{3/2}}
 -\frac{e^{it \cos k_F} \left(n\sin nk_F+\cot k_F \cos nk_F\right)}{\pi t^2 \sin^2 k_F}+O(t^{-5/2}).
\end{equation}
Then we are left with the problem of finding the asymptotics expansion of $C(k_F,n,t)$. It is convenient to split it
into two parts $C=C_1+C_2$ where $C_{1}$ and $C_2$ contain the imaginary and real part of the logarithm respectively. We have
\begin{equation}
C_1(k_F, n,t)=ik_FA(n,t)-i\pi B(k_F,n,t) 
\end{equation}
and the asymptotic follows from \eqref{S_as_a}-\eqref{S_as_B}, whereas
\begin{equation}
\label{S_c2}
 C_2(k_F,n,t)=\frac{1}{4\pi}\int_{-\pi}^{\pi} dk\, e^{it\cos k}e^{ink} \log \left(\frac{1-\cos [k_F-k]}{1-\cos[k_F+ k]}\right).
\end{equation}
The asymptotics is dominated by two ordinary stationary points at $k_s=0,\pi$ where the phase vanishes; they produce
terms $O(t^{-3/2})$ since the integrand is vanishing there. To understand the contribution of the logarithmic singularity at
$k=\pm k_F$ we rewrite the integral as
\begin{equation}
\label{Sc2_int}
C_2(k_F,n,t)=\frac{i}{2\pi}\int_{0}^{\pi} dk\, e^{it\cos k}\sin(nk)\log \left(\frac{1-\cos [k_F-k]}{1-\cos[k_F+ k]}\right) 
\end{equation}
and notice that we can expand the logarithm for $k\rightarrow k_F$ as $2\log(k-k_F)+\text{reg.}$. When crossing
the logarithmic singularity the imaginary part jumps by $2i\pi$ and generates an effective boundary contribution at $k_F$
that can be evaluated by integration by parts. One finds the final result
\begin{align}
\label{S_as_C2}
\nonumber
C_2(k_F,n,t)= &e^{it\cos k_F}\left( -\frac{i\sin nk_F}{t\sin k_F}+\frac{\cot k_F \sin nk_F-n\cos n k_F}{t^2\sin^2k_F}\right)+\\
&\sqrt{\frac{2}{\pi}}\frac{n\, i^{n+1}}{(\sin k_F)t^{3/2}}\left[\cos k_F \sin \left(\frac{\pi}{4}+\frac{n\pi}{2}-t\right)+i\sin \left(\frac{\pi}{4}-\frac{n\pi}{2}+t\right)\right]+O(t^{-5/2})
\end{align}
The asymptotics expansions \eqref{S_as_a}, \eqref{S_as_B} and \eqref{S_as_C2} solve our problem as the derivative of any correlator
can be expressed in terms of them. We decompose the two-point function into the sum of four pieces $r,l$ (right/left)
and $T,H$ (Toeplitz/Hankel) according to the function $f^{r/l}_{T/H}(k,q)$ entering in each of them, see (\ref{SflT}-\ref{SfrH}),
and get
\begin{align}
\nonumber
\boxed{\partial_t\langle c^{\dagger}_n(t)c_m(t)\rangle_{H,l}}&=-\frac{1}{4\pi}[-(A(-n,-t)C(k_F^l,1-m,t)-A(m,t)C(k_F^l,1-n,-t))\\
\label{mess1}
+&A(1-n,-t)C(k_F^l,2-m,t)-A(m-1,t)C(k_F^l,2-n,-t)],
\end{align}
\begin{align}
\nonumber
&\boxed{\partial_t\langle c^{\dagger}_n(t)c_m(t)\rangle_{T,l}}=
\frac{1}{4\pi}[C(k_F^l,-n+1,-t)A(m,t)-A(-n+1,-t)C(k_F^l,m,t)-C(k_F^l,-n,-t)A(m-1,t)\\
\label{mess2}
&+A(-n,-t)C(k_F^l,m-1,t)-2\pi i(A(-n+1,-t)B(k_F^l,m,t)-A(-n,-t)B(k_F^l,m-1,t))],
\end{align}
\begin{align}
\nonumber
\boxed{\partial_t\langle c^{\dagger}_n(t)c_m(t)\rangle_{H,r}}=&
-\frac{1}{4\pi}[-(A(m,t)C(k_F^r,1+n,-t)-A(-n,-t)C(k_F^r,1+m,t))+\\
\label{mess3}
&A(m-1,t)C(k_F^r,n,-t)-A(-n+1,-t)C(k_F^r,m,t)],
\end{align}
\begin{align}
\nonumber
\boxed{\partial_t\langle c^{\dagger}_n(t)c_m(t)\rangle_{T,r}}=&
\frac{1}{4\pi}[C(k_F^l,-n+1,-t)A(m,t)-A(-n+1,-t)C(k_F^l,m,t)-C(k_F^l,-n,-t)A(m-1,t)\\
\label{mess4}
&+A(-n,-t)C(k_F^l,m-1,t)+2\pi i(B(k_F^r,-n+1,-t)A(m,t)-B(k_F^r,-n,-t)A(m-1,t))].
\end{align}
Plugging \eqref{S_as_a}, \eqref{S_as_B} and \eqref{S_as_C2} into (\ref{mess1}-\ref{mess2}) we
determine the full asymptotics expansion of the time-derivative of the two-point functions. The result can be integrated back to
derive the approach to the NESS of all the correlators. We gave a relevant example for $n=0$ and $m=1$, in the main text Eq.
\eqref{asymptotic_J}. To validate the method we also provide the asymptotic expansion of the real part of the
correlator $\mathcal{G}_{-3}(t)$ at $k_F^l=\pi-k_F^r=2\pi/3$.
It reads
\begin{multline}
\label{S_sp34}
\Re[\mathcal{G}_{-3}(t)]\stackrel{t\gg 1 }{\simeq}
 \frac{\sqrt{3}}{14\pi}+\frac{-8/\sqrt{3}+\pi}{2\pi^2 t}-\frac{7\cos(2t)}{4\pi t^2}\\
 -\frac{2\cos\bigl[\frac{1}{6}(\pi-9 t)\bigr]-3(3+\sqrt{3})\cos(t/2)+\cos(3t/2)+
 [-9+2\sqrt{3}-2\sqrt{3}\cos(t)]\sin(t/2)}{6(\pi t)^{3/2}}.
\end{multline}
The comparison between the asymptotic formula \eqref{S_sp34} and the exact curve obtained by the power series expansion, discussed
in the previous section, is shown in the main text.  
\hypertarget{sec:semiclassics}{\appsubsection{Semiclassical limit}}
\begin{figure}[t]
\centering
\includegraphics[width=6cm]{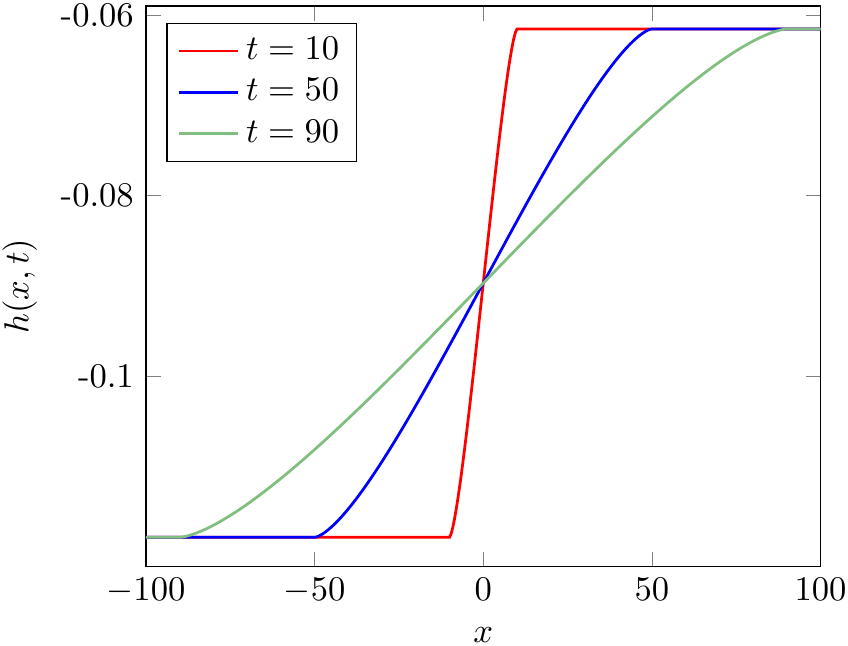}
\caption{Semiclassical limit for the energy density profile $h(x,t)$ plotted at different times $t$ and for $\beta_l=1$, $\beta_r=0.5$.
The function \eqref{energy_density} is expected to exactly reproduce the large $x$, $t$ limit of the energy density
in a fermionic chain, whose halves are at time $t=0$ thermalized independently at temperatures $T_{l/r}=\beta_{l/r}^{-1}$.}
\label{fig_energy}
\end{figure}
We now obtain results for the correlators in the semiclassical limit.
The stationary points of the phases in \eqref{Fourier} of the main text
are the solutions of the following equations ($v(k)\equiv d\varepsilon(k)/dk$)
\begin{align}
\label{stat_eq1}
 & v(k_s)-x/t=0, \\
 \label{stat_eq2}
 & v(q_s)-y/t=0,
\end{align}
We restrict ourselves to the case $(x-y)/t\rightarrow 0$;  taking the difference of (\ref{stat_eq1}-\ref{stat_eq2}) we observe
that stationary points either coincide $q_s=k_s$ or are shifted by $\pi$, $k_s=\pi-q_s$.
Let us introduce new variables $Q,K,d,X$ through
\begin{align}
 & k=K+Q/2,\quad q=K-Q/2,\\
 & x=X-d/2,\quad y=X+d/2.
 \end{align}
When the saddle points coalesce the integrand in \eqref{Fourier} is singular and linearization around $Q_s=k_s-q_s=0$ gives the leading
behavior in the semiclassical region. We obtain 
\begin{equation}
\label{semiclassic_CF}
 \langle c^{\dagger}_x (t) c_y(t)\rangle\stackrel{\text{semiclassics}}{\longrightarrow}
 \frac{1}{2\pi}\int_{-k_F^l}^{k_F^l}dK~e^{iKd}\Theta(-X+v(K)t)+\frac{1}{2\pi}\int_{-k_F^r}^{k_F^r}dK~e^{iKd}\Theta(X-v(K)t),
\end{equation}
where we have used the standard representation of the Heaviside step function
$\Theta(x)=\frac{1}{2\pi i}\int_{\mathbb R}dQ~\frac{e^{iQ x}}{Q-i0}$.  For $X/t\rightarrow 0$ we are back with correlation functions in the
translation invariant NESS. After elementary integration one finds
\begin{equation}
 \label{S_NESS}
\langle c^{\dagger}_x c_y\rangle_{\rm NESS}=\frac{1}{2\pi i}\frac{e^{ik_F^r(x-y)}-e^{-ik_F^l(x-y)}}{x-y},
\end{equation}
the particle momenta in the NESS are obtained shifting all particle momenta in the ground state $k\in[-k_0,k_0]$ by a
constant amount $\delta=k_F^l-k_0$ ($k_0=\pi/2$),  see \eqref{SNESS}.

As we discussed briefly in the main text, semiclassical results can be obtained in other physically relevant contexts.
For example, when the two halves of the chain have same densities but different inverse temperatures $\beta_l$ and $\beta_r$
one expects, after the quench, a ballistic energy transport with
a inhomogeneous energy density profile $h(x,t)$. In the semiclassical limit,
the function $h(x,t)$ is again a scaling function $h(x/t)$ that can be determined analytically.
Let $\rho_0$ be the initial density matrix for the two disconnected chains then~\cite{DMV14}
\begin{equation}
\label{therm_f}
 f(k,q)=\Tr[\rho_0 c^{\dagger}(k)c(q)]=
 -\frac{n_r(k)+n_r(q)}{4\pi(1-e^{i(q-k-i0)})}+\frac{n_l(k)+n_l(q)}{4\pi(1-e^{i(q-k+i0)})}+\text{regular},
\end{equation}
where we have omitted regular terms for $k\rightarrow q$ and $n_{l/r}(k)=1/[1+e^{\beta_{l/r}\varepsilon(k)}]$.
Notice that for cosine dispersion relation the choice of a zero chemical potential fixes the particle density to
be homogeneous and equal $1/2$, therefore particle transport is absent. The similarity between
\eqref{therm_f} and \eqref{SflT}, \eqref{SfrT} is evident. Proceeding in complete analogy with the calculation
of the correlation functions, explained few lines before, we obtain for the energy density profile
\begin{equation}
\label{energy_density}
 h(x,t)\stackrel{\text{semiclassics}}{\longrightarrow}\frac{1}{2\pi}\int_{-\pi}^{\pi}dK~\varepsilon(K)n_l(K)\Theta(-x+v(K)t)
 +\frac{1}{2\pi}\int_{-\pi}^{\pi}dK~\varepsilon(K) n_r(K)\Theta(x-v(K)t),
\end{equation}
where we assumed $\varepsilon(k)=-\cos k$. A plot is given in Fig. \ref{fig_energy}. Nothing prevents a study of the
large-deviation function of the energy-flow, for example with $\beta_l=0$ and $\beta_r=\infty$, in the same spirit
of \cite{EislerRacz_PRL2013}. Those aspects will be not investigated here.
\hypertarget{sec:semiclassicsbis}{\appsubsection{An alternative derivation of the semiclassical limit}}
We provide here another derivation of our main result, that is closer to the standard stationary phase procedure. As already emphasized the main complication is that the stationary points in $k$ and $q$ almost coincide, and the denominator is singular in such a limit. This difficulty may be circumvented by using the same derivative trick as in section~\hyperlink{sec:corrections}{C1}.

\begin{figure}[htbp]
\centering
\includegraphics[width=12cm]{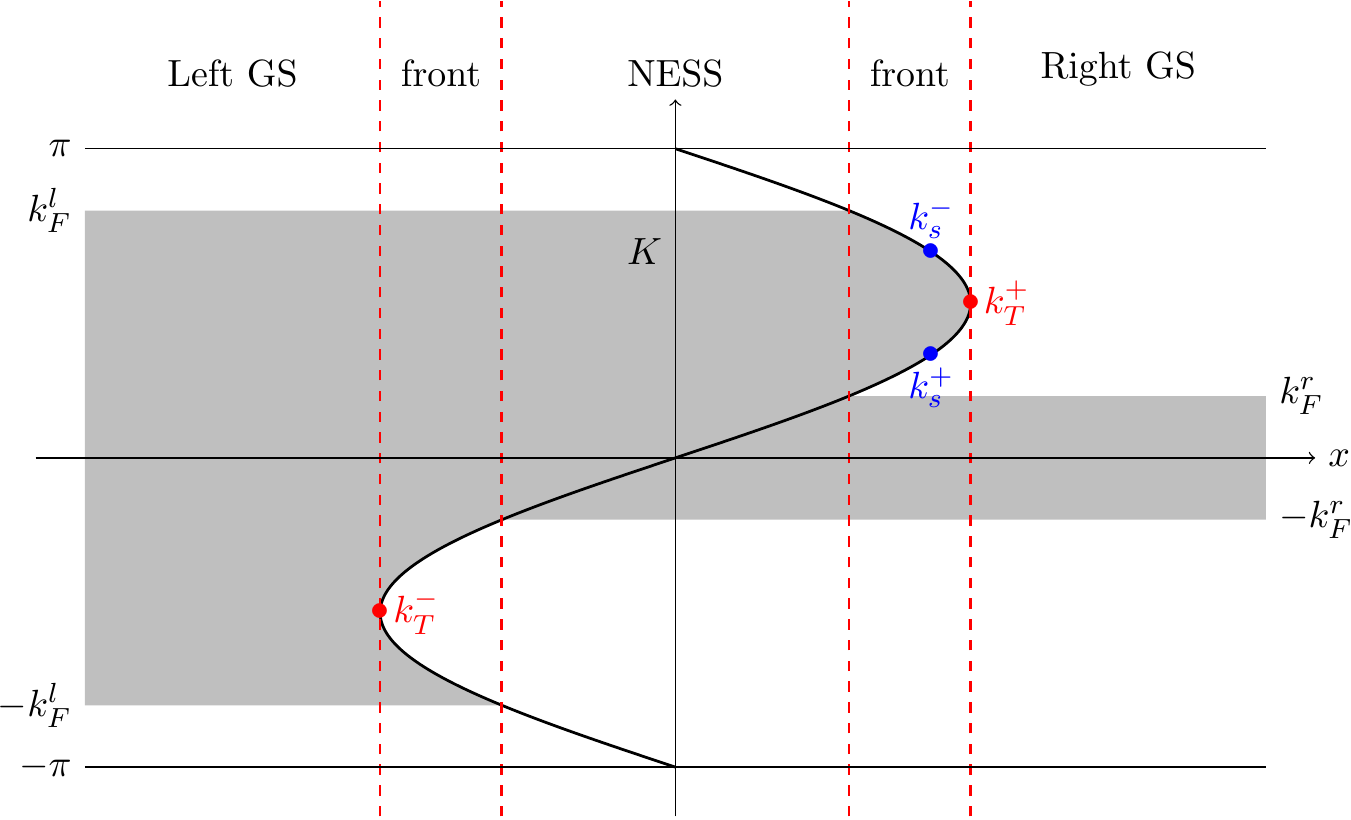}
\caption{The picture shows the integration domain in the variable $K$ constrained by the
two Heaviside step functions $\Theta(-x+v(K)t)$ and $\Theta(x-v(K)t)$ in \eqref{semiclassic_CF},
for $x=y$ and fixed time $t$. The black curve $K(x)$ is the solution of the stationary phase equation $v(K)=x/t$.
The stationary points (Eq.~\eqref{statk}) in the inhomogeneous front region are shown in blue.
Approaching the right edge of the front
$k_s^+, k_s^-\rightarrow k_T^+=\pi/2$, the turning point. Near the turning point,
\textit{i.e.} for $x/t\sim 1$, correlations have non-trivial
subleading corrections with respect to their value in the bulk of the right ground state; these corrections are described by the Airy
kernel~\eqref{corr_edge}.}
\label{fig_semi}
\end{figure}

Let us explain the method on the term involving $f_T^{l}(k,q)$. Since $f_H^{l}(k,q)$ is regular, it will not contribute in the scaling limit. For the same reason, only the pole in $f_T^{l}(k,q)$ matters. Therefore, to the leading order we have to evaluate
\begin{equation}
\label{corr_left}
 \braket{c_x^\dag(t)c_y(t)}_{l}=\int_{-\pi}^{\pi} \frac{dk}{2\pi}\int_{-k_F^l}^{k_F^l}\frac{dq}{2\pi} \frac{e^{i[\varepsilon(k)-\varepsilon(q)]t-ik x+iqy}}{1-e^{i(q-k+i0^+)}}.
\end{equation}
Now we consider the time-derivative of the previous equation. We obtain
\begin{equation}
 \frac{\partial}{\partial t} \braket{c_x^\dag(t)c_y(t)}_{l}=\int_{-\pi}^{\pi} \frac{dk}{2\pi}\int_{-k_F^l}^{k_F^l}\frac{dq}{2\pi} \frac{i[\varepsilon(k)-\varepsilon(q)]}{1-e^{i(q-k)}}
 e^{i\Phi_x(k)-i\Phi_y(q)},
\end{equation}
which is now regular at $k=q$. The phase $\Phi_x(k)$ is given by
\begin{equation}
 \Phi_x(k)=\varepsilon(k)t-kx.
\end{equation}
We now have to look for the points where the phase is stationary. These are solution of the equation
\begin{equation}
 v(k_s)=\frac{x}{t}
\end{equation}
where $v(k)=\frac{d\varepsilon(k)}{dk}=\sin k$. For $|x/t|<1$ there are two real solutions
\begin{equation}
\label{statk}
 k_s^+=\arcsin \frac{x}{t}\qquad,\qquad k_s^{-}=\pi\sign\, x-\arcsin \frac{x}{t}.
\end{equation}
The treatment for $\Phi_y(q)$ is similar, and we get
\begin{equation}
\label{statq}
 q_s^+=\arcsin \frac{y}{t}\qquad,\qquad q_s^{-}=\pi\sign\, y-\arcsin \frac{y}{t}.
\end{equation}
Let us now assume that the $k_s^{\pm},q_s^\pm$ belong to the integration domain, which means we are in the inhomogeneous region $-1<x/t,y/t<-\sin k_F^l$. Around the stationary points $\Phi_x'(k_s^\pm)=0$, the phase may be approximated by
\begin{equation}
 \Phi_x(k)=\Phi_x(k_s^\pm)+\frac{1}{2}\Phi_x^{\prime\prime}(k_s^{\pm})(k-k_s^\pm)^2,
\end{equation}
and the same goes for $\Phi_y(q)$. There are in principle four contributions at $(k_s^+,q_s^+)$, $(k_s^-,q_s^-)$, $(k_s^+,q_s^-)$, $(k_s^-,q_s^+)$. One can check that the last two give fast oscillating contributions, which become subleading when integrated back. Neglecting those two we obtain
\begin{equation}
\label{stat_phase_der}
 \frac{\partial}{\partial t} \braket{c_x^\dag(t)c_y(t)}_{l}=\frac{1}{2\pi}\frac{1}{\sqrt{\Phi_x''(k_s^+)}}\frac{1}{\sqrt{\Phi_y''(q_s^+)}}\frac{i[\varepsilon(k_s^+)-\varepsilon(q_s^+)]}{1-e^{i(q_s^+-k_s^+)}}e^{i[\Phi_x(k_s^+)-\Phi_y(q_s^+)]}\;+\;\Big\{k_s^+\to k_s^-\;;\;q_s^+\to q_s^-\Big\}.
\end{equation}
At distance $|x-y|/t\ll 1$ we have $k_s^+=q_s^+$ (resp. $k_s^-=q_s^-$) to the leading order. Therefore,
\begin{align}
  \frac{\partial}{\partial t} \braket{c_x^\dag(t)c_y(t)}_{l}&=\frac{v(k_s^+)e^{i(y-x)k_s^+}}{2\pi \sqrt{\Phi_x''(k_s^+)}\sqrt{\Phi_y''(k_s^+)}}\;+\;\Big\{k_s^+\to k_s^-\Big\}\\
  &=\frac{\sin(k_s^+)\left[e^{i(y-x)k_s^+}+e^{i(y-x)k_s^-}\right]}{2\pi t \cos(k_s^+)}
\end{align}
Noticing that $\tan k_s^+=t\frac{\partial k_s^+}{\partial t}$ , this may be rewritten as
\begin{align}
  \frac{\partial}{\partial t} \braket{c_x^\dag(t)c_y(t)}_{l}&=\frac{\partial k_s^+}{\partial t}\frac{e^{i(y-x)k_s^+}}{2\pi}-\frac{\partial k_s^-}{\partial t}\frac{e^{i(y-x)k_s^-}}{2\pi}\\\label{eq:result}
  &=\frac{\partial}{\partial t}\left[\int_{k_s^-}^{k_s^+}\frac{dK}{2\pi}e^{-iK(x-y)}\right],
\end{align}
which is the expected result in the left inhomogeneous (front) region $-1\leq x/t,y/t \leq -\sin k_F^l$,
see Fig.~\ref{fig_semi}. Note however that the method only gives the correlations up to some integration constant, which depends on $x$ and $y$. This constant may be fixed by using a continuity argument. Indeed by applying the same method, we find that the derivative vanishes in the regions $|x/t|>1$ and $|x/t|<\sin k_F^l$. Since we know that far on the left (resp. right) the initial correlations are that of the initial left (resp. right) ground state, we use the boundary condition at $t=|x|$ to fix the integration constant in (\ref{eq:result}). We then do the same at $|x/t|=\sin k_F^l$ to obtain the correlations in the region $|x/t|<\sin k_F^l$. 
\hypertarget{sec:Airy}{\appsubsection{Correlations near the edge and the Airy kernel}}
Let us focus for simplicity on the the right edge of the front and assume $k_F^l>\pi/2$. At the leading order correlations are given by their ground-state value in the right-part
of the chain and are in particular time-independent. However,
approaching the boundary of the light-cone the stationary points  $k_s^+$ and $q_s^+$
in (\ref{statk}-\ref{statq}) both reach the turning point $k_T^+=\pi/2$ (see Fig.~\ref{fig_semi}) where
the second derivative  of the phases in \eqref{stat_phase_der} vanishes. As a consequence one finds
non-trivial subleading corrections that can be computed evaluating the integral
\begin{equation}
\label{corr_sing}
\langle c^{\dagger}_x(t) c_y(t)\rangle_{l}=\int_{-\pi}^{\pi}\frac{dk}{2\pi}
\int_{-k_F^l}^{k_F^l}\frac{dq}{2\pi}\frac{e^{-i \cos kt -ixk+i\cos q t+iyq}}
{1-e^{i(q-k+i0)}},
\end{equation}
when $k, q\rightarrow k_T^{+}$.  Expanding  up to third-order in the
phase,  we get
\begin{equation}
\label{correlator_edge}
\langle c^{\dagger}_x(t) c_y(t)\rangle_{l}\stackrel{\text{\tiny{near the edge}}}{\longrightarrow} e^{-ik_T^+(x-y)}\int_{-\infty}^{\infty}\frac{d\tilde k}{2\pi}
\int_{-\infty}^{\infty}\frac{d\tilde q}{2\pi}\frac{e^{-i\tilde{k}(x-t)-i\frac{t\tilde k^3}{3!}+i\tilde{q}(y-t)
+i\frac{t\tilde q^3}{3!}}}
{i(\tilde{k}-\tilde{q}-i0)}, 
\end{equation}
with $\tilde k=k-k_T^+$ and $\tilde q=q-k_T^+$.
Introducing the scaling variables $X=(x-t)\left(\frac{2}{t}\right)^{1/3}$,
$Y=(y-t)\left(\frac{2}{t}\right)^{1/3}$ and defining $K=(t/2)^{1/3} \tilde k$, $Q=(t/2)^{1/3} \tilde q$ we can
rewrite \eqref{correlator_edge} as
\begin{align}
\nonumber
 \langle c^{\dagger}_x(t) c_y(t)\rangle_{l}&\stackrel{\text{\tiny{near the edge}}}{\longrightarrow} 2^{1/3}t^{-1/3}e^{-ik_T^+(x-y)}
 \int_{-\infty}^{\infty}\frac{dK}{2\pi}\int_{-\infty}^{\infty}\frac{dQ}{2\pi}
 \frac{e^{-iKX-iK^3/3+iQY+iQ^3/3}}{i(K-Q-i0)}\\
 \label{corr_edge}
 &=2^{1/3}t^{-1/3}e^{-ik_T^+(x-y)} K(X,Y),
\end{align}
where $K(X,Y)$ is the celebrated Airy kernel~\cite{TW93}
\begin{equation}
\label{Airy_ker}
K(X,Y)=\frac{\text{Ai}(X)\text{Ai}'(Y)-\text{Ai}'(X)\text{Ai}(Y)}{X-Y}.
\end{equation}
The last passage in \eqref{corr_edge} follows from the relation
$-(\partial_X+\partial_Y)K(X,Y)=\text{Ai}(X)\text{Ai}(Y)$ and the integral representation of the Airy
function $\text{Ai}(X)=\int_{\mathbb R}\frac{dK}{2\pi} e^{iXK+iK^3/3}$. Extension of the result
to finite temperature correlation
functions (see the end of Sec. \hyperlink{sec:semiclassics}{C2}) is also possible.

\clearpage

\begin{figure}[htbp]
\centering
\includegraphics[width=12cm]{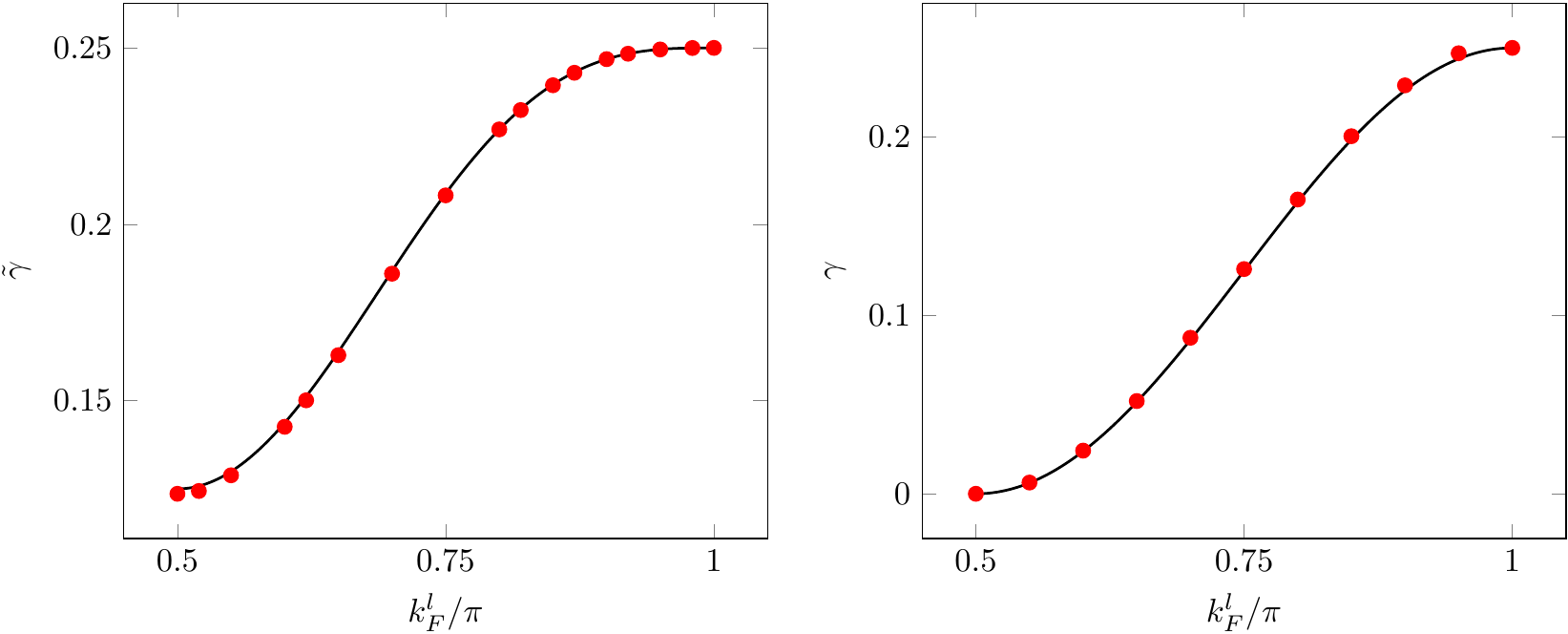}
\caption{\textit{Left.} We show numerical data for the coefficient $\tilde{\gamma}$ (red dots) compared with the
exact formula \eqref{tildegamma}. \textit{Right.} The red dots show the coefficient $\gamma$ for the large-time limit of
the Loschmidt echo extracted by the numerics. The black line is the empirical formula $\gamma(x)=\frac{1}{4}\cos^2(x\pi)$.}
\label{fig_LE_sup}
\end{figure}

\pagebreak
\hypertarget{sec:Loschmidt}{\appsection{Loschmidt echo}}
This section is devoted to the study of the Loschmidt echo
after the quench from the domain-wall initial state.
We provide two separate derivations (Sec. \hyperlink{sec:le2}{D2}-\hyperlink{sec:le3}{D3}) of the result
\begin{equation}\label{eq:someleresult}
 \mathcal{L}(t)=e^{-t^2/4}
\end{equation}
quoted in the main text, as well as a generalization to any dispersion relation.
Before doing so, we first briefly discuss the short- and long-time behavior of the Loschmidt
echo for arbitrary fillings.
\hypertarget{sec:le1}{\appsubsection{Short- and long- time behavior}}
For convenience we start from
the definition $\mathcal{L}(t)=|\langle\psi_0|e^{-iHt}e^{iH_0 t}|\psi_0\rangle|^2$, where $H_0$ is the pre-quench Hamiltonian. If
$|\psi_0\rangle$ is an eigenstate of $H_0$, this is clearly immaterial. By expanding for small $t$ one obtains
$\mathcal{L}(t)=1-\tilde{\gamma}t^2+o(t^2)$,
with
 \begin{equation}
 \tilde{\gamma}=\text{Var}(H-H_0)\equiv\langle\psi_0|(H-H_0)^2|\psi_0\rangle-\bigl[\langle\psi_0|(H-H_0)|\psi_0\rangle\bigr]^2.
\end{equation}
If we have a quadratic operator $A$ which is represented by the matrix $\mathcal A$ in the fermion real Fock space then
the variance of such operator in the initial state $|\psi_0\rangle$ is
\begin{equation}
\label{variance}
 \text{Var}(A)=\text{Tr}[C^T \mathcal{A} \tilde C \mathcal{A}],
\end{equation}
where $\tilde C=\mathbf{1}-C$ and $C$ is the initial state correlation matrix. For the filling fraction quench the matrix $C$
is given in the main text Eq. \eqref{cor_zero}. It is possible to check that the values of $\tilde{\gamma}$ are independent from the
cut-off $L$, used to represent the matrices $H$, $H_0$ and $C$ and we find the exact expression
\begin{equation}
\label{tildegamma}
 \tilde{\gamma}(x)=\frac{2 \pi^2 + 4 x^2 + 2 \pi \sin 2 x + \sin^2 (2 x) - 
 4 x (\pi + \sin 2 x)}{8\pi^2},
\end{equation}
where $x=k_F^l$. A numerical check of formula \eqref{tildegamma} is presented in Fig. \ref{fig_LE_sup}.

As already commented in the
main text the Loschmidt echo large time behavior is unfortunately accessible only numerically. The data are compatible with a
gaussian decay $\mathcal{L}(t)\stackrel{t\gg 1}{\rightarrow} e^{-\gamma t^2}$, with coefficient $\gamma$ well reproduced by the empirical
formula $\gamma(k_F^l)=\frac{1}{4}\cos^2(k_F^l)$, see again Fig. \ref{fig_LE_sup}.   

\hypertarget{sec:le2}{\appsubsection{Domain wall limit: determinant derivation}}
The first strategy is to consider a finite system of size $L$, and to express the Loschmidt echo as a determinant. This determinant is then evaluated in the limit $L\to \infty$. 
The Hamiltonian can be put in diagonal form
\begin{equation}
 H=\sum_q \varepsilon_q d_q^\dag d_q,
\end{equation}
with
\begin{equation}
 d_q^\dag =\sqrt{\frac{2}{L+1}}\sum_{j=1}^{L}\sin \frac{q\pi j}{L+1}c_j^\dag.
\end{equation}
Here $q\in \{1,2,\ldots,L\}$ and $\varepsilon_q=\cos \frac{q\pi }{L+1}$. More general dispersion relations will be discussed later. Note that we chose to shift the labels of the sites by $L/2$, so that the open chain now starts at site $1$ and ends at site $L$.  With our convention the initial state is $\ket{\psi_0}=\prod_{j=1}^{L/2}c_j^\dag \ket{0}$. In a Heisenberg picture, the real space fermion operators are given by
\begin{equation}\label{eq:discretebessel}
 c_j^\dag(t)=e^{i H t}c_j^\dag e^{-i H t}=\sqrt{\frac{2}{L+1}}\sum_{q}\sin \frac{q\pi j}{L+1}e^{i \varepsilon_q t}d_q^\dag =\frac{2}{L+1}\sum_{l=1}^{L}\left(\sum_{q=1}^L \sin\frac{q\pi j}{L+1} \sin\frac{q\pi l}{L+1}e^{i t \varepsilon_q}c_l^\dag\right).
\end{equation}
In the limit $L\to \infty$ for fixed $t$, the sum over $q$ in the previous equation converges to a sum of two Bessel functions, so that
\begin{equation}
 c_j(t)=\sum_{l=1}^{\infty} \left(J_{j-l}(t)-J_{j+l}(t)\right)c_l^\dag,
\end{equation}
where $J_x(t)=\int_{-\pi}^{\pi}\frac{dq}{2\pi}e^{it \cos q-iq x}$ is a Bessel function of the first kind. Note the appearance of $J_{j+l}(t)$, which comes from the boundary. Using Wick's theorem, the echo may be expressed as a determinant. 
In the limit $L\to \infty$, we obtain
\begin{equation}\label{eq:ledet}
 \Braket{e^{ i H t}}=\lim_{L\to \infty}\left[\det_{1\leq j,l\leq L/2}\left(J_{j-l}(t)-J_{j+l}(t)\right)\right].
\end{equation}
It is important to understand that we have first taken the limit $L\to \infty$ to get the Bessel functions from (\ref{eq:discretebessel}), and \emph{only} then taken the infinite determinant limit. Hence the calculation is not mathematically rigorous. The physical justification of this assumption is the presence of a light-cone, which ensures for any finite time $t$ that the effects of the boundaries at distance $L/2\to\infty$ are suppressed.  

The r.h.s of the previous equation is the determinant of a matrix whose elements only depends on $j-l$ and $j+l$. Such determinants are called Toeplitz ($j-l$) + Hankel ($j+l$) determinants. They have been widely studied in the mathematical
literature, starting from the work of Szeg\H{o} \cite{Szegobis}. The important object to consider for such evaluations is the \emph{symbol} of the determinant, namely the inverse Fourier transform of $J_x(t)$. Here the symbol simply follows from the definition of the Bessel function, and is
\begin{equation}\label{eq:simplesymbol}
 g(k)=\sum_{x\in \mathbb{Z}}J_x(t)e^{-ikx}=e^{i t \cos k}\qquad,\qquad k\in [-\pi,\pi].
\end{equation}
We note $[g(k)]_x=\int_{-\pi}^{\pi}\frac{dk}{2\pi}g(k)e^{ixk}=J_x(t)$ the corresponding Fourier coefficients. 
With this at hand, the asymptotics of (\ref{eq:ledet}) follow from a theorem derived in Ref.[\onlinecite{Basorbis}]. For sufficiently regular symbols, the following asymptotic formula holds
\begin{equation}
 \lim_{n\to \infty}\det_{1\leq j,l\leq n}\left([g(k)]_{j-l}-[g(k)]_{j+l}\right)=\exp\left(\frac{1}{2}\sum_{x=1}^\infty x\left([\log g(k)]_x\right)^2-\sum_x [\log g]_{2x}\right),
\end{equation}
provided $g(-k)=g(k)$ and $[\log g(k)]_0=0$, which is the case here. The (logarithm of the) symbol (\ref{eq:simplesymbol}) corresponding to the Bessel function has only one non zero Fourier coefficient, so that we get
\begin{equation}
 \Braket{e^{i H t}}=e^{-t^2/8},
\end{equation}
which is exact at all times in an infinite system. Squaring this gives the claimed result (\ref{eq:someleresult}) for the Loschmidt echo. 

It is important to emphasize that the choice of boundary conditions is not innocent in such calculations. In case of periodic boundary conditions, there are two light cones developing after the quench, one at $x=L/2,L/2+1$, but also one at $x=0,L$. Because of these two light cones, the Lochmidt echo in a periodic system is the square of the one in an open system. This observation serves as an alternative and simpler way to compute it. Indeed with PBC we get a Toeplitz determinant
\begin{eqnarray}\label{eq:limit1}
 \Braket{e^{i tH_{\rm per}}}&=&\lim_{n\to \infty}\det_{1\leq j,l\leq n}\left([g(k)]_{j-l}\right)\\\label{eq:limit2}
 &=&\exp\left(\sum_{x=1}^\infty x[\log g(k)]_x[\log g(k)]_{-x}\right)\\
 &=&e^{-t^2/4}\\
 &=& \Braket{e^{i tH_{\rm open}}}^2.
\end{eqnarray}
To go from (\ref{eq:limit1}) to (\ref{eq:limit2}) we have use the strong Szeg\H{o} limit theorem \cite{Szegobis}.
Generalization to other dispersion relations is also possible. Let us consider an Hamiltonian with longer-range hoppings
\begin{equation}
 H=\sum_{j\in \mathbb{Z}}\sum_{\alpha=1}^{p}\left( u_\alpha c_{j+\alpha}^\dag c_j+h.c\right).
\end{equation}
The range $p$ of the longest hopping term need not be finite, but let us assume that for now. Then, regularizing the Hamiltonian on a periodic ring with $L$ sites, the dispersion relation is
\begin{equation}
 \varepsilon(k)=2\sum_{\alpha=1}^p u_\alpha \cos (\alpha k),
\end{equation}
so that the symbol of the Toeplitz determinant becomes
\begin{equation}
 g(k)=e^{2i t\sum_{\alpha=1}^p u_\alpha \cos (\alpha k)}.
\end{equation}
Application of the Szeg\H{o} limit theorem then gives
\begin{equation}
 \Braket{e^{it H_{\rm per}}}=e^{-\left(\sum_{\alpha=1}^p \alpha u_\alpha^2\right)t^2}.
\end{equation}
Taking the square-root we obtain
\begin{equation}
 \Braket{e^{it H}}=\exp\left(-\frac{1}{2}\left[\sum_{\alpha=1}^p \alpha \,u_\alpha^2\right]t^2\right),
\end{equation}
and so
\begin{equation}
 \mathcal{L}(t)=\exp\left(-\left[\sum_{\alpha=1}^p \alpha \,u_\alpha^2\right]t^2\right).
\end{equation}
We have assumed that $p$ is finite in the calculation, but the result should also hold for an infinite number of Fourier coefficients, provided the series $\sum_{\alpha\geq 1} \alpha u_\alpha^2$ converges.

\hypertarget{sec:le3}{\appsubsection{Domain wall limit: combinatorial derivation}}
We provide an alternative derivation that uses only combinatorial means. Let us focus on the dispersion relation $\varepsilon(k)=u\cos k$ for now. The method is similar in spirit to the short-time expansion presented in the main text, and can be made fully rigorous. The real-space Hamiltonian is
\begin{equation}
 H=u\sum_{j \in \mathbb{Z}} \left(c_{j+1}^\dag c_j +c_{j}^\dag c_{j+1}\right).
\end{equation}
Expanding $e^{i H t}$ in power series yields
\begin{equation}
 \Braket{e^{i H t}}=\sum_{m=0}^{\infty} \frac{(i t)^m}{m!} \Braket{H^m},
\end{equation}
where $\braket{.}$ denotes the average in the domain wall initial state. 
Therefore, evaluating the Loschmidt echo boils down to finding an exact expression for all the moments $\braket{H^m}$ of the Hamiltonian. 
The initial state can be pictured as shown below:
\vspace{0.5cm}
\begin{center}
\begin{tikzpicture}
\draw[thick] (-6,0) -- (5,0);
\draw[dashed,thick] (-6.5,0) -- (-6,0);
\draw[dashed,thick] (5,0) -- (5.5,0);
 \foreach \x in {-6,-5,-4,-3,-2,-1}{
 \filldraw (\x,0) circle (2.5pt);
 \draw[very thick] (-1-\x,0) circle (2.5pt);
 \pgfmathparse{int(\x+1)}\let\j\pgfmathresult;
 \draw (\x,-0.5) node {\j};
 }
 \foreach \x in {0,1,2,3,4,5}{
 \pgfmathparse{int(\x+1)}\let\j\pgfmathresult;
 \draw (\x,-0.5) node {\j};
 }
 \draw (-0.5,-0.4) -- (-0.5,0.4);
\end{tikzpicture}
\end{center}
\vspace{0.5cm}
The sites on the left (resp. right) are filled (resp. empty). 
The action of $H$ on $\ket{\psi_0}=\prod_{x \leq 0} c_x^\dag \ket{0}$ obeys rather simple rules, which we explain now. 
Each particle tries to go to one of it's nearest neighbor site, provided there is not already a particle here. At the first time step the only possibility is for the rightmost particle to go one step to the right. Hence 
\begin{equation}
 H \ket{\psi_0}=H \ket{\ldots 111111|000000\ldots}=u\ket{\ldots 111110|100000\ldots}.
\end{equation}
Here a $1$ represents a site occupied by a fermion, and a $0$ an empty site (hole). To improve readability a vertical bar $|$ between sites $0$ and $1$ was added. $H^2$ is given by
\begin{eqnarray}
 H^2 \ket{\psi_0}&=&H u\ket{\ldots 111110|100000\ldots} \\
 &=& u^2\ket{\ldots 111111|000000\ldots}+u^2\ket{\ldots 111110|010000\ldots}+u^2\ket{\ldots 111101|10000\ldots}.
\end{eqnarray}
Since $\braket{H^m}=\braket{\psi_0|H^m|\psi_0}$ selects the initial state in this expansion, we have $\braket{H}=0$ and $\braket{H^2}=u^2$. Said differently, $\braket{H^m}$ counts the number of ways for hardcore particles to move to the left and to the right, where at each step only one particle moves, with the constraint that all particles have to come back to their initial positions after $m$ steps. Because a particle is forced to move at each step, it is easy to see that $\Braket{H^{2m+1}}=0$ for integer $m$. With these rules at hand, obtaining $\braket{H^{2m}}$ boils down to a known combinatorial problem. The result is
\begin{equation}\label{eq:simple}
 \Braket{H^{2m}}=u^{2m}(2m-1)!!,
\end{equation}
where $(2m-1)!!=1.3.5\ldots (2m-1)$ is the double factorial. This formula is remarkably simple, but its derivation is not. Proving it can be done by (i) reformulating the problem in terms of oscillating Young tableaux, and then (ii) using a known bijection between those and perfect matchings. 

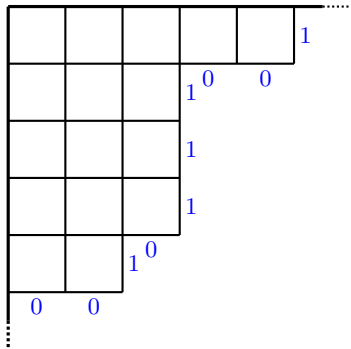
\begin{figure}[htbp]
 \begin{tikzpicture}[scale=0.76]
  \draw[very thick] (0,0) -- (0,-5.5);\draw[very thick,densely dotted] (0,-5.5) --(0,-6);\draw[thick] (1,0) -- (1,-5);\draw[thick] (2,0) -- (2,-5);
  \draw[thick] (3,0) -- (3,-4);\draw[thick] (4,0) -- (4,-1);\draw[thick] (5,0) -- (5,-1);
  \draw[very thick] (0,0) -- (5.5,0);\draw[thick,densely dotted] (5.5,0) --(6,0);
  \draw[thick] (0,-1) -- (5,-1);\draw[thick] (0,-2) -- (3,-2);\draw[thick] (0,-3) -- (3,-3);\draw[thick] (0,-4) -- (3,-4);\draw[thick] (0,-5) -- (2,-5);
  \draw[blue] (0.5,-5.25) node {$0$};\draw[blue] (1.5,-5.25) node {$0$};\draw[blue] (2.5,-4.25) node {$0$};\draw[blue] (3.5,-1.25) node {$0$};\draw[blue] (4.5,-1.25) node {$0$};
  \draw[blue] (2.2,-4.5) node {$1$};\draw[blue] (3.2,-3.5) node {$1$};\draw[blue] (3.2,-2.5) node {$1$};\draw[blue] (3.2,-1.5) node {$1$};\draw[blue] (5.2,-0.5) node {$1$}; 

 \end{tikzpicture}
\caption{Young tableau corresponding to the state $\ket{\psi}=\ket{\ldots1111100101|1100100000\ldots}$. The relevant particle configuration after removal of the spurious ones and zeroes is $00101|11001$. The initial state corresponds to the empty diagram.}
\label{fig:tableaumapping}
\end{figure}

The first step (i) goes as follows~\cite{Leeuwen}. To each particle configuration may be associated a Young tableau, i.e. a collection of left-justified boxes, with non increasing row height. This is done by first removing all ones part of a semi-infinite sequence of ones, and doing the same for the zeroes. The height of the first row is then the number of remaining ones (which equals the number of remaining zeroes). We then draw an horizontal edge to the right (resp. vertical edge up) for each zero (resp. one) encountered. The initial state $\ket{\psi_0}$ corresponds to an empty diagram. This procedure is illustrated in Fig.~\ref{fig:tableaumapping}, on the example $\ket{\psi}=\ket{\ldots1111100101|1100100000\ldots}$. Moving a particle then corresponds to adding or removing a unit box. 

Evaluating $\braket{H^{2m}}$ amounts to counting sequences of young tableaux that start and end at the empty diagram, with $m$ steps up and $m$ steps down. Such sequences are called oscillating tableaux, and they have been shown to be in bijection with perfect matchings \cite{Roby}. Since the number of perfect matchings of $(1,2,\ldots,2m)$ is exactly $(2m-1)!!$, the result of Ref.~\cite{Roby} immediately implies (\ref{eq:simple}). Hence
\begin{align}
 \Braket{e^{i H t}}&=1+\sum_{m=1}^{\infty} \frac{(itu)^{2m}}{(2m)!}(2m-1)!!\\
 &=e^{-\frac{1}{2}u^2 t^2}.
\end{align}
Squaring this and setting $u=1/2$, we finally obtain
\begin{equation}
 \mathcal{L}(t)=e^{-t^2/4},
\end{equation}
as claimed in the main text. The technique can also be generalized to other dispersion relations. In that case the longer-range movements of the particles correspond to the addition of bigger boxes, called ribbons, with signed weights that depend on the number of particles overtaken at each step. This exactly encodes the free fermionic nature of the problem. Such oscillating ribbon tableaux have been studied in Ref.~\cite{DGBPN} by similar bijection techniques. Using the bijection described in the reference, it then follows that \cite{Kfootnote}
\begin{equation}\label{eq:generalizedmoments}
 \Braket{H^{2m}}=(2m-1)!!\left[u_1^2+2u_2^2+3u_3^2+\ldots\right],
\end{equation}
which implies
\begin{equation}
  \mathcal{L}(t)=\exp\left(-\left[\sum_{\alpha\geq 1} \alpha \,u_\alpha^2\right]t^2\right).
\end{equation}
This result coincides with the determinant method. Note that the Loschmidt echo is zero at all times $t>0$ in case the sum $\sum_{\alpha\geq 1} \alpha u_\alpha^2$ diverges.

\end{document}